# Quantum imaging of biological organisms through spatial and polarization entanglement

Yide Zhang[†], Zhe He[†], Xin Tong[†], David C. Garrett, Rui Cao, and Lihong V. Wang[*]

*Caltech Optical Imaging Laboratory, Andrew and Peggy Cherng Department of Medical Engineering, Department of Electrical Engineering, California Institute of Technology, 1200 E. California Blvd., MC 138-78, Pasadena, CA 91125, USA*

† These authors contributed equally.

* Correspondence should be addressed to L.V.W. (LVW@caltech.edu).

## Abstract

Quantum imaging can potentially provide certain advantages over classical imaging. Thus far, however, the signal-to-noise ratios (SNRs) are poor; the resolvable pixel counts are low; biological organisms have not been imaged; birefringence has not been quantified. Here, we introduce quantum imaging by coincidence from entanglement (ICE). Utilizing spatially and polarization entangled photon pairs, ICE exhibits higher SNRs, greater resolvable pixel counts, imaging of biological organisms, and ghost birefringence quantification; it also enables 25 times greater suppression of stray light than classical imaging. ICE can potentially empower quantum imaging towards new applications in life sciences and remote sensing.

## Introduction

Since van Leeuwenhoek's first microscope, optical imaging has been widely used to noninvasively investigate the structures and dynamics of various physical and biological systems[1,2]. The key advantage of optical imaging is that the interaction of non-ionizing light with molecules provides rich molecular information about biological samples. Aided by the convenience and compactness of optical systems, optical imaging has served as the workhorse for biological researchers and medical practitioners behind a wide variety of discoveries[3]. In the past two decades, advanced optical imaging techniques have been developed to allow super-resolution[1,4] and high-speed[5,6] bioimaging. However, to achieve high resolution and high imaging speed, most optical imaging techniques require intense illumination that can disrupt or damage the biological processes under investigation[2]. Low-intensity illumination may lead to a low signal-to-noise ratio (SNR) due to shot noise and stray light.

Recently, to overcome the limitations of existing optical imaging techniques that rely on classical light sources, quantum imaging approaches that use correlated, entangled, or squeezed photons have been developed[7–11]. Compared with classical optical imaging, quantum imaging has the following advantages[12]. First, the classical shot-noise limit can be broken, allowing for sub-shot-noise (SSN) imaging under low-intensity illumination[11,13–17]. Second, stray light can be suppressed[10,18,19]. Third, super-resolution imaging beyond the diffraction limit can be enabled[8,20–25]. Empowered by these advantages, quantum imaging has been employed to investigate biological specimens[8,11,26], which have complex structures and may be susceptible to photobleaching and thermal damage. Despite the advantages, quantum images of biological specimens reported to date still suffer a low SNR because (1) the conditions required to achieve SSN are stringent[13,15–17,27] and (2) the SNRs in most quantum imaging approaches are low[10,12,19]. Moreover, existing quantum imaging approaches usually have low resolvable pixel counts (i.e., the ratios of the field of view (FOV) to the spatial resolution)[7–11] and thus are not suited for practical biological studies, which often demand systematic investigation of multiple parts in a biological system with an FOV across a whole organism. Finally, quantum imaging techniques so far only measure transmittance (absorption) or phase contrast, whereas classical techniques support additional contrast such as birefringence[1,2,28].

Here we present ICE, a higher-SNR, greater-resolvable-pixel-count, and birefringence-sensitive quantum imaging technique that generates high-quality images of biological specimens. Under low-intensity illumination, ICE employs a new SSN algorithm that utilizes the covariance of the raw images to achieve a higher SNR than the classical counterpart. Concurrently, ICE substantially increases the SNR over existing quantum imaging techniques by accommodating multiple spatial modes of the entangled photon pairs in each pixel, where a single spatial mode is constrained by the diffraction limit of the system[29,30]. The spatial resolution of ICE is determined by both the signal and idler photons through a quantum effect named “entanglement pinhole”. In the entanglement pinhole effect, when an entangled photon pair is captured concurrently by two detectors, one detector functions non-classically as a pinhole on the object being imaged by the other detector. Further, ICE increases the resolvable pixel counts indefinitely through raster scanning and is 25 times more resilient to stray light than classical imaging. Consequently, ICE enables quantum imaging of whole organ (mouse brain) slices and organisms (zebrafish) with an FOV of up to 7 mm × 4 mm, and can be operated in the presence of ambient lighting, thus suitable for practical biological studies. Finally, ICE exploits the polarization entanglement of the photon pairs for ghost birefringence imaging, where the birefringence properties of an object can be remotely and instantly measured without changing the polarization states of the photons incident on the object. The quantum advantages of ICE, therefore, enable the observation of biological specimens under conditions that cannot be satisfied with classical imaging, as well as the remote sensing of birefringence.

## Results

### Sub-shot-noise quantum imaging using multi-mode entangled photons

In ICE (Fig. 1, details in Methods), we use two β-barium borate (BBO) nonlinear crystals with perpendicularly aligned optical axes to produce hyperentangled photon pairs, which are simultaneously entangled in spatial mode, polarization, and energy[31,32], through the type-I spontaneous parametric down-conversion (SPDC) process. Most quantum imaging techniques reported to date evenly distribute the spatial modes of entangled photons across multi-pixel cameras[10,13,19,33,34], leading to a small number of spatial modes per pixel, a low coincidence rate, and, consequently, a low SNR in the image. In comparison, ICE increases the coincidence rate and SNR of quantum images by directly focusing the multi-mode SPDC beam onto the object, resulting in substantially more spatial modes in each pixel. We record the signal ($N_s$), idler ($N_i$), and coincidence ($N_c$) counts from the two single-photon counting modules (SPCMs) while raster scanning the object through the focused SPDC beam to image the transmittance of the object. Whereas $N_s$ and $N_c$ provide classical and quantum (ICE) images of the object, respectively, $N_i$ can further improve the SNR of the images through SSN signal retrieval using our covariance-over-variance (CoV) algorithm (Supplementary Note 1, Supplementary Figs. 1 and 2). Compared with state-of-the-art SSN methods such as ratio and optimized subtraction[15,17,27], our CoV algorithm achieves higher SNRs for both classical ($N_s$) and quantum ($N_c$) imaging, as demonstrated through simulations (Supplementary Fig. 3) and experiments (Supplementary Fig. 4).

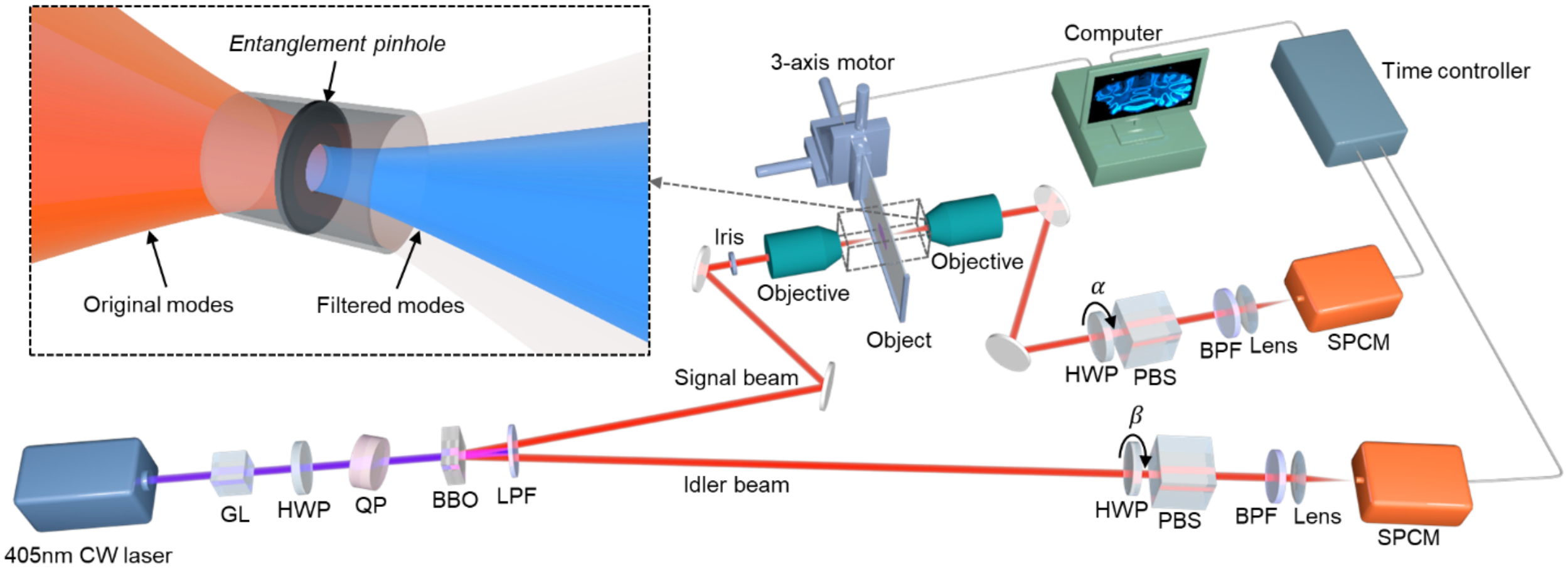

**Fig. 1 Experimental setup.**

Setup schematics. CW, continuous wave; GL, Glan–Laser polarizer; HWP, half-wave plate; QP, quartz plate; BBO, β-barium borate crystals; LPF, long-pass filter; PBS, polarizing beam splitter; BPF, band-pass filter; SPCM, single-photon counting module. Inset, illustration of the entanglement pinhole.

Despite the higher coincidence rate and SNR, acquiring images by raster scanning a multi-mode beam is generally undesired in classical imaging, as the multi-mode beam leads to a broad point spread function (PSF) and consequently a poor spatial resolution. However, as shown in the inset of Fig. 1, the spatially entangled photon pairs in ICE enable a quantum effect named "entangled pinhole," where the detector in the idler arm functions as a pinhole on the object in the signal arm (Supplementary Note 2, Supplementary Fig. 5). Relying on the true coincidences from spatially entangled photons (Supplementary Note 3, Supplementary Fig. 6), the entanglement pinhole filters out a portion of the spatial modes in the SPDC beam (Supplementary Fig. 7) and improves spatial resolution and depth of field (DOF) over classical imaging. As shown in Fig. 2a, the classical image of a US Air Force (USAF) resolution target can only resolve groups 4 and 5, whereas ICE can clearly resolve groups 6 and 7. Further, ICE maintains higher resolution over a long axial distance (Fig. 2b). To quantify the resolution and DOF experimentally, we acquired the edge spread functions (ESFs) of the images at different $z$ positions. We then computed the line spread functions (LSFs) and their full width at half maximum (FWHM) to estimate the spatial resolutions (Methods). As shown in Fig. 2c, ICE has finer resolution than classical imaging from $z = -0.3$ mm to $z = 0$ mm. To calculate the DOF of the system, we repeated the same resolution analysis with a finer step size (10 µm) through approximately 700 µm along the *z*-axis (Fig. 2d). To align the foci, the curve for ICE has been shifted to the right by 43 µm. By fitting the experimental data, the focal resolutions of classical imaging and ICE are determined to be 14.4 ± 0.6 µm and 10.4 ± 0.4 µm, respectively, demonstrating that ICE improves the resolution by 38% over classical imaging; the DOFs, on the other hand, are determined to be 92 ± 2 µm and 95 ± 2 µm for classical imaging and ICE, respectively.

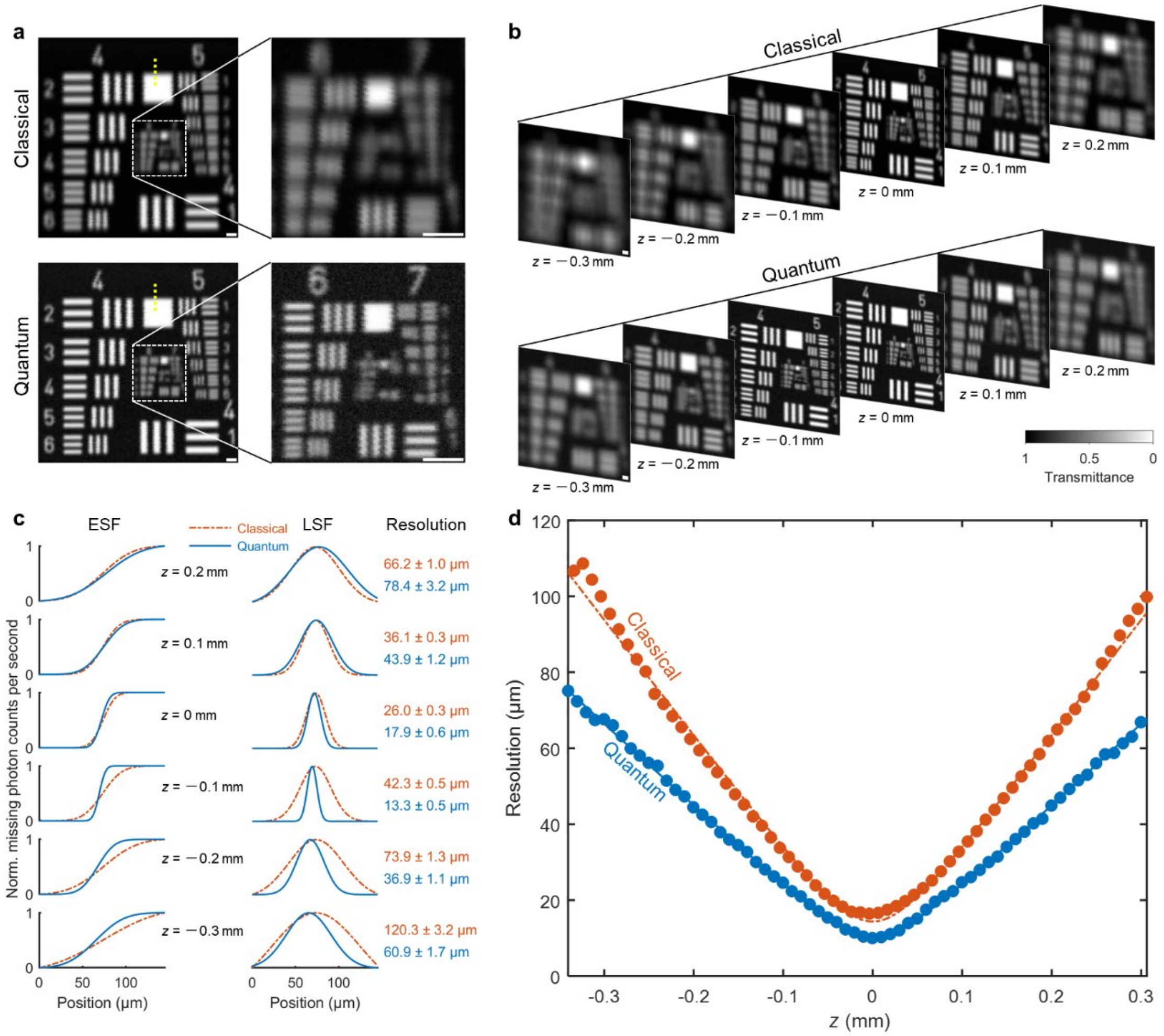

**Fig. 2 Effect of the entanglement pinhole on ICE.**

**a**,**b**, Classical imaging and ICE of a USAF resolution target at focus (**a**) and at different *z* positions (**b**), where $z = 0$ mm denotes the focus of classical imaging. **c**, Edge spread functions (ESFs), line spread functions (LSFs), and spatial resolutions measured at different *z* positions. The ESFs were fitted from the profiles of the yellow dotted lines in **a**. The means and standard errors of the resolution are shown on the right. **d**, Resolution versus *z* for classical imaging and ICE. Dots represent experimental measurements. Solid and dash-dotted lines denote fits. Norm., normalized. Scale bars, 50 μm.

Compared with existing quantum imaging techniques that have been typically demonstrated with thin biological samples[7–11] (e.g., < 10 μm), ICE provides a larger DOF, thus enabling the observation of thick objects. Here, we imaged 500-μm thick agarose with randomly embedded carbon fibers of 6 μm diameter each. As shown in Supplementary Fig. 8a, ICE can resolve the carbon fibers better than classical imaging throughout an axial range of 300 μm. The profiles along the yellow dashed lines demonstrate ICE's ability to resolve three closely located fibers that cannot be clearly distinguished classically. ICE has imaged all targets in the object more clearly due to both the higher spatial resolution and the large DOF over the classical counterpart. Specifically, comparing the averages of the 3D stacks acquired classically and through ICE (Supplementary Fig. 8b), one can see that, within a 3D volume of 1000 × 1000 × 300 $\mu m^3$, the carbon fibers in the ICE stack are clearly better resolved and have sharper edges than those in the classical stack.

**Quantum imaging of biological organisms in the presence of stray light**

By raster scanning the object, ICE provides an FOV that can be extended indefinitely. We imaged a slice of a whole organ (the cerebellum of a mouse brain) with a 7 mm × 4 mm FOV, whose anatomical structures are annotated in Supplementary Fig. 9a. The ICE image (Supplementary Fig. 9b) outperforms the classical counterpart (Supplementary Fig. 9a) with a higher resolution, as seen in the two regions of interest (ROIs) in Supplementary Figs. 9c and e. Compared with the line profiles of the classical images (Supplementary Figs. 9d and f), the narrower trenches and peaks in the ICE profiles confirm an improved resolution across the large FOV.

In addition to the large FOV, ICE also demonstrates robust stray light resistance due to coincidence detection. To quantify ICE's resilience to ambient lighting, an LED was added to the system to introduce stray light (Supplementary Fig. 10). We acquired classical and ICE images of a biological organism, i.e., an agarose-embedded zebrafish, in a 3.5 mm × 2.3 mm FOV while the LED was randomly turned on and off to simulate randomly fluctuating ambient light (Fig. 3). The zebrafish was positioned such that its torso was oblique to the imaging plane (Supplementary Fig. 11). As shown in Fig. 3a, while the classical imaging is severely degraded by the stray light, ICE is almost unaffected. We further quantify the robustness of ICE to stray light by acquiring a series of images of carbon fibers under different stray light intensities (Fig. 3b). Using the images acquired without the stray light as the ground truth, we calculated the structural similarity index measure (SSIM) of each image to quantify the degradation of the image quality due to stray light[35] (Methods). The SSIM ranges from 0 to 1, where higher values indicate less degradation. The SSIM versus the stray light optical power is plotted in Fig. 3c. In accordance with the images in Fig. 3b, the classical images degrade quickly with an LED optical power above 0.1 mW, while ICE maintains a high SSIM even with an LED optical power above 1 mW. To simplify the comparison, we use an order-of-magnitude degradation (SSIM = 0.1) as a threshold to find the corresponding LED optical powers, found as 0.18 mW and 4.41 mW for the classical imaging and ICE, respectively. Therefore, ICE suppresses stray light 25 times more effectively than classical imaging. The advantage of ICE can also be seen in the difference between the two SSIM curves, i.e., ΔSSIM, shown in Fig. 3c. This advantage of ICE is attributed to coincidence detection, which is disturbed only by accidental coincidence counts. Despite its sufficient intensity to degrade a classical image, stray light acts as an uncorrelated source, causing negligible coincidence counts.

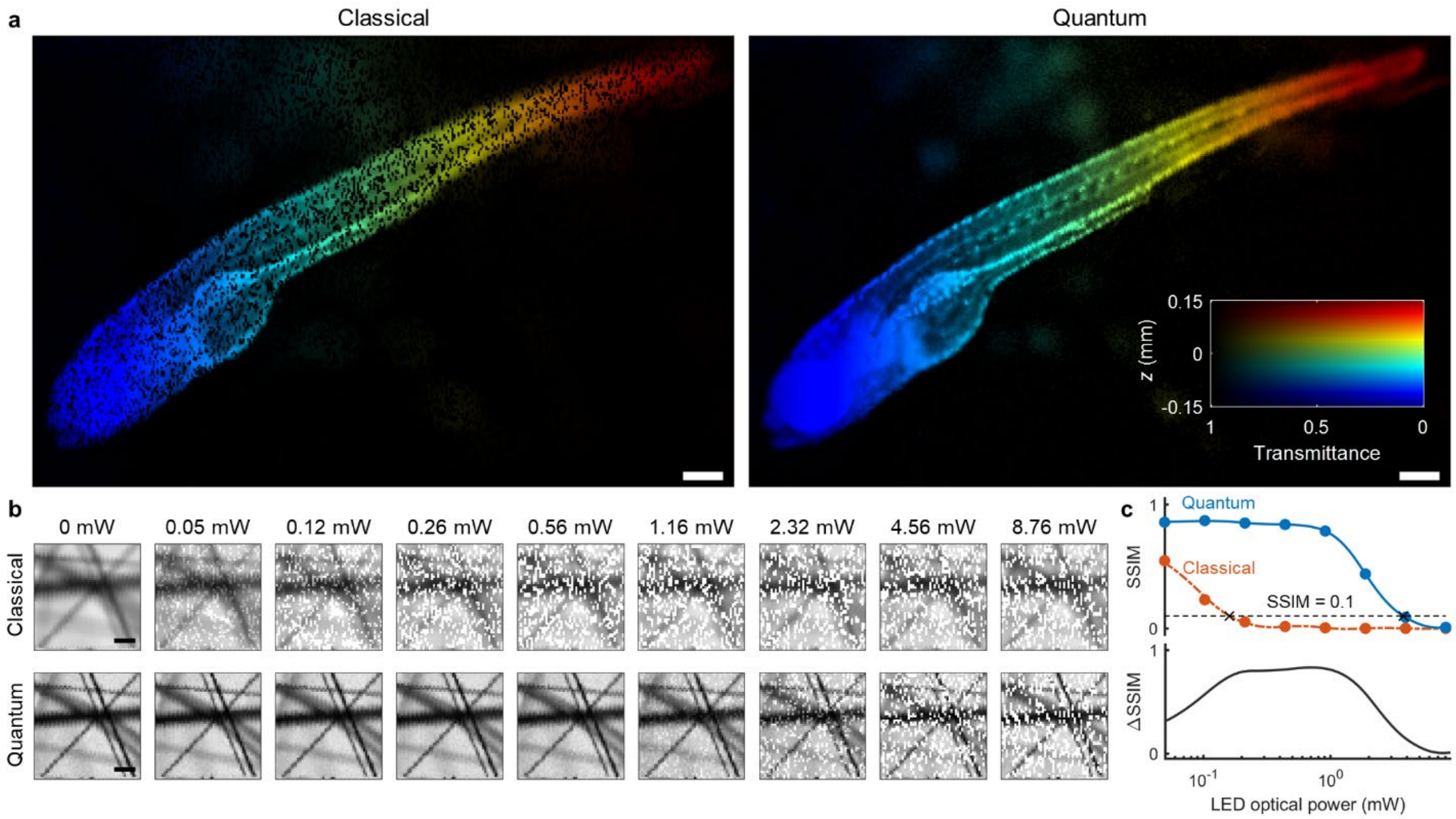

**Fig. 3 ICE in the presence of stray light.**

**a**, Classical and ICE images of a whole zebrafish in the presence of stray light. The pseudo colors encode the $z$ positions of the sample. Scale bars, 200 μm. **b**, Classical and ICE images of carbon fibers acquired at different stray light optical powers. Scale bars, 100 μm. **c**, Top, structural similarity index measure (SSIM) calculated between the images in **b** and the ones without the stray light. Black dashed line, a threshold (SSIM = 0.1) used to quantify the robustness of ICE and classical imaging. Bottom, difference between the SSIM curves for ICE and classical imaging.

**Ghost birefringence imaging through polarization entanglement**

Whereas most existing quantum imaging techniques rely on the spatial entanglement of SPDC photon pairs[7,13,17,19], quantum imaging modalities utilizing polarization entanglement, such as the quantum holography[10], have been developed recently. The polarization entanglement of the SPDC photon pairs in our system can be characterized by Bell's test[36,37] (Supplementary Note 4, Supplementary Fig. 12). With an $S$ value of $2.78 \pm 0.01 > 2$, our system shows a substantial violation of the Clauser–Horne–Shimony–Holt (CHSH) inequality[38], demonstrating strong polarization entanglement[32]. By using hyperentangled photon pairs that are simultaneously entangled in spatial mode and polarization[31,32], ICE measures the birefringence properties of an object without changing the polarization states of the photons incident on the object.

We evaluated the ghost birefringence imaging capability of ICE by imaging a biological organism—a whole zebrafish embedded in agarose. We kept the polarization of the signal photons incident on the object constant ($\alpha = 0°$) while changing the polarization angles of the idler photons, which do not traverse the object, to four different angles ($\beta = 0°, 45°, 90°, 135°$). Whereas the four classical images exhibited little differences (Supplementary Fig. 13), the ICE images were substantially modulated by the birefringence properties of the zebrafish (Fig. 4a). Following the theory in Supplementary Note 5, the four ICE images could be used to calculate the transmittance, the angle of the principal refractive index (Fig. 4b), and the phase retardation between the two refractive index axes (Fig. 4c) of the zebrafish, providing additional biologically relevant information that has not been obtained with existing quantum imaging techniques. Furthermore, because of polarization entanglement, measuring the idler photon's polarization state instantly determines the incident signal photon's, thus allowing instant measurement of the object's birefringence properties, regardless of its distance. With the capability to remotely and instantly measure the birefringence properties of an object by changing the polarization states of the photons that do not probe the object, ICE can be used in remote sensing applications where the source is too far to be controlled in real time (Supplementary Fig. 14).

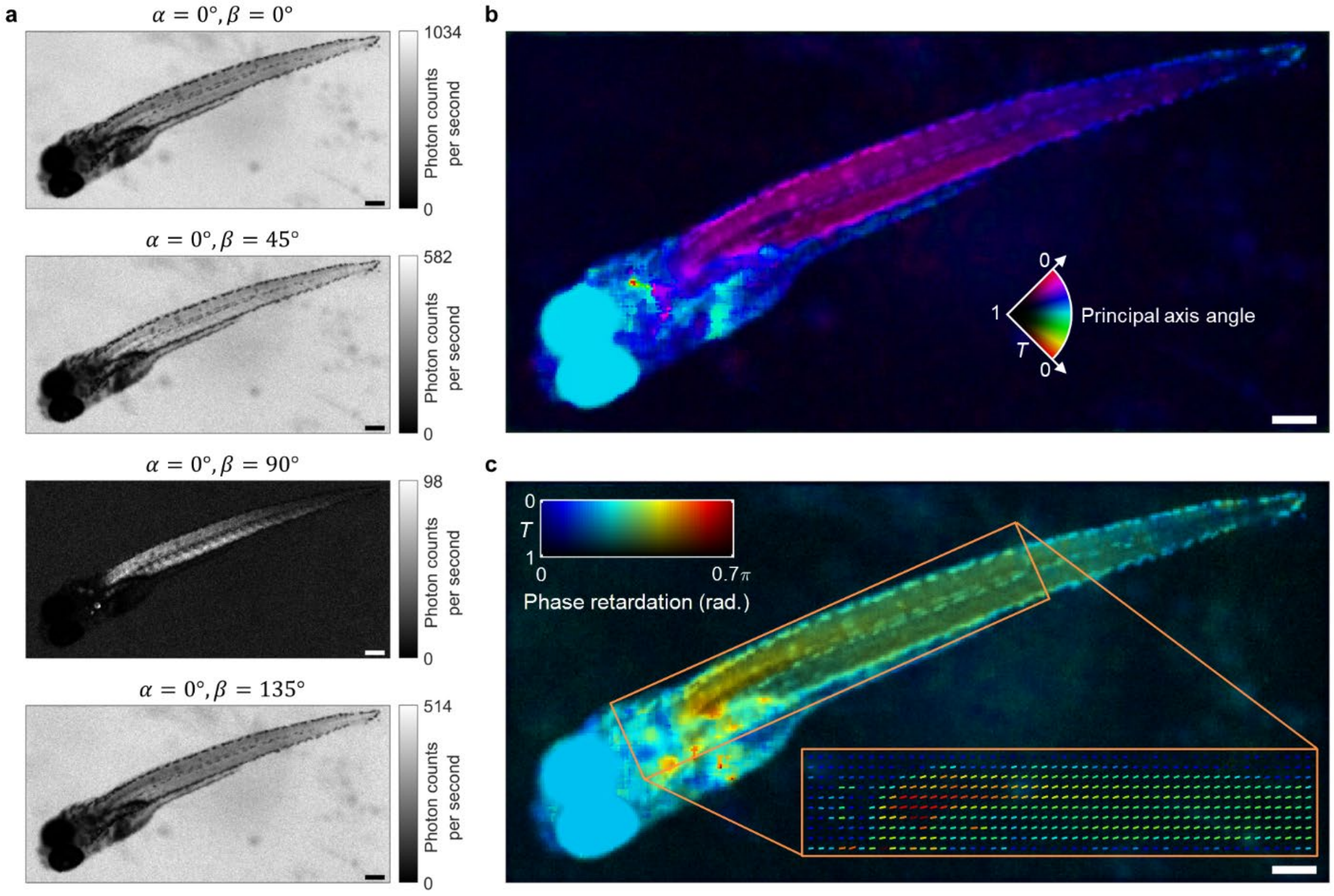

**Fig. 4 Ghost birefringence imaging of a whole zebrafish with ICE.**

**a**, ICE images acquired with a polarizer of a constant angle $\alpha$ and a polarizer of a variable angle $\beta$. **b**, Transmittance (*T*) and principal refractive index axis angle (pseudo colors) calculated using the ICE images in **a**. **c**, Transmittance (*T*) and phase retardation between the two refractive index axes (lines and pseudo colors) calculated using the ICE images in **a**. Scale bars, 200 µm.

## Discussion

Although imaging by coincidence can be achieved with a classical source[39], the SNR of the image will be substantially lower compared to that of ICE under the same illumination intensity (Supplementary Note 6, Supplementary Figs.15 and 16), and the advantages enabled by spatial and polarization entanglement, such as SSN performance and ghost birefringence imaging, will be unavailable. We also note that, despite the similarity in using spatially entangled photon pairs and detecting coincidence for imaging, ICE fundamentally differs from ghost imaging (GI)[40] or correlation plenoptic imaging (CPI)[41] for the following reasons: (1) ICE generates a direct image of the object through raster scanning over a theoretically unlimited FOV, whereas GI and CPI provide an indirect, ghost image of the object through triggering a multi-pixel camera with a limited FOV; (2) The signal arm of ICE contributes to spatial resolution, whereas the signal arms of GI and CPI do not; (3) ICE images substantially more spatial modes per pixel than GI and CPI; (4) ICE exploits polarization entanglement in addition to the spatial entanglement used in GI and CPI (see Supplementary Note 7 and Supplementary Fig. 17 for detailed comparison).

Despite the advantages, ICE has the following limitations. First, the pixel dwell time is currently 1 s, limited by the low SPDC efficiency of the BBO crystal[42]. Second, due to the utilization of multi-mode SPDC beams, ICE has a lower spatial resolution compared to the Abbe limit of resolution[1,2]. These problems could be solved in the future by using a more powerful quantum source[42]. A strong entangled photon source with high coincidence rates could substantially improve the imaging speed, and the SPDC beam could be filtered to a single spatial mode for diffraction-limited imaging while maintaining a sufficient SNR. Third, the entanglement pinhole is a virtual pinhole that filters SPDC modes in coincidence detection. In practice, all the SPDC photons in the signal arm still transmit through the object, which undergoes an illumination intensity higher than the two-photon coincidence used for quantum imaging. Nevertheless, the photon flux of all the SPDC photons on the object is less than 20 kHz (Supplementary Fig. 16), which equals $4.9 \times 10^{-15}$ W, an ultralow illumination intensity that is safe for photosensitive biological specimens.

To conclude, we have experimentally demonstrated ICE using hyperentangled photon pairs, achieving high-quality quantum bioimaging with higher SNR, greater resolvable pixel counts, and ghost birefringence quantification. As showcased using the thick biological organism (whole zebrafish) and the whole organ (mouse brain) slice with an FOV substantially larger than those of existing quantum images (Supplementary Fig. 18, Supplementary Table 1), these features allow for systematic observations in complex biological specimens. Rather than competing with classical imaging, ICE offers complementary benefits and additional opportunities such as SSN performance, stray light resistance, and ghost birefringence quantification. With these benefits and opportunities, ICE is expected to find more applications in life sciences where low illumination intensity, ambient lighting, or precise measurements are required, and in remote sensing where the source cannot be controlled in real time.

# Methods

## Experimental setup

In our system (Fig. 1), a paired set of BBO crystals ($5 \times 5 \times 0.5$ mm$^3$ each, PABBO5050-405(I)-HA3, Newlight Photonics) was cut for type-I spontaneous parametric down-conversion (SPDC) at 405 nm wavelength. The two crystals were mounted back-to-back with one crystal rotated by 90 degrees about the normal axis to the incidence surface. The pump was a 405 nm continuous wave laser (LM-405-PLR-40-4K, Coherent) with an output power of 40 mW. A Glan–Laser polarizer (GL10-A, Thorlabs) and a half-wave plate (WPA03-H-405, Newlight Photonics) were used to adjust the pump laser beam to be linearly polarized at 45 degrees relative to the vertical axis. A quartz plate (QAT25100-A, Newlight Photonics) tilted about its vertically oriented optical axis was used to pre-compensate for the phase difference between the horizontal and vertical polarization components of the SPDC photons. The pump laser beam then passed through the BBO crystals and generated a ring of SPDC photons with a half opening angle of 3 degrees. A long-pass filter with a cut-on wavelength of 715 nm (LWPF1030-RG715, Newlight Photonics) was used to block the pump beam after the crystals. While the SPDC idler beam was directly sent to a half-wave plate (WPA03-H-810, Newlight Photonics) for polarization selection, the signal beam, whose size was adjusted by an iris (ID20, Thorlabs), was focused via an objective lens (LI-20X, 0.4 NA; LI-10X, 0.25 NA; LI-4X, 0.1 NA; Newport) onto a microscope slide. The microscope slide was mounted on a 3-axis motor (462-XYZ-M, each axis installed with an LTA-HS motorized actuator, Newport). The transmitted SPDC signal beam was collected by another objective lens of the same type and then sent to another half-wave plate (WPA03-H-810, Newlight Photonics) for polarization selection. The two half-wave plates were mounted on two motorized precision rotation mounts (PRM1Z8, Thorlabs), each followed by a polarizing beam splitter (PBS201, Thorlabs), an $810 \pm 30$ nm band-pass filter (NBF810-30, Newlight Photonics), a collection lens (LA1131, Thorlabs), and a single-photon counting module (SPCM-AQRH-16, Excelitas Technologies). The two SPCMs were connected to a time controller (ID900-TCSPC-HR, ID Quantique) with a digital time resolution of 13 ps to measure both raw singles photon counts and coincidence counts. The time controller and the 3-axis motor were synchronized and controlled by a computer. While motor-scanning the microscope slide holding the object, the raw singles counts of the SPDC signal beam and the coincidence counts of the signal and idler beams were used to form the classical and ICE images of the object, respectively. The pixel dwell time was 1 s. The whole setup was covered by a light-shielding box.

**Characterization of polarization entanglement**

The entanglement of the SPDC signal and idler photon pairs was evaluated using Bell's test with the Clauser–Horne–Shimony–Holt (CHSH) inequality (Supplementary Note 4)[36,37]. Denoting the angles of the half-wave plates on the signal and idler paths as $\alpha$ and $\beta$, respectively, we recorded the coincidence counts $N(\alpha,\beta)$ at each step with an acquisition time of 1 s and a coincidence detection window of 8 ns. The correlation value was calculated by

$$E(\alpha,\beta)=\frac{N(\alpha,\beta)+N(\alpha+90°,\beta+90°)-N(\alpha+90°,\beta)-N(\alpha,\beta+90°)}{N(\alpha,\beta)+N(\alpha+90°,\beta+90°)+N(\alpha+90°,\beta)+N(\alpha,\beta+90°)}.$$

The CHSH inequality was then evaluated at the angle pairs $\alpha \in \{0°, 45°\}$ and $\beta \in \{22.5°, 67.5°\}$ based on the value of

$$S=|E(0°,22.5°)-E(0°,67.5°)|+|E(45°,22.5°)+E(45°,67.5°)|.$$

As shown in Supplementary Fig. 12, our system shows a strong violation of the CHSH inequity with $S=2.78\pm0.01>2$ estimated by calculating the mean and standard error of $S$ values measured from 10 rounds of Bell's tests.

**Sample preparation**

Four types of objects have been imaged. The wild-type zebrafish was fixed by 4% paraformaldehyde (PFA) solution five days post fertilization. After fixation, the zebrafish was washed 3-4 times using PBS in a fume hood prior to agarose embedding. The agarose-embedded zebrafish was mounted onto a glass slide and sealed with a coverslip to prevent dehydration during the experiment. To prepare the brain slice, a brain was obtained from a Swiss Webster mouse (Hsd: ND4, Harlan Laboratories) and fixed in 3.7% paraformaldehyde solution at room temperature for 24 h. After paraffin embedding, coronal sections (10 µm thick) of the brain were cut. Standard hematoxylin and eosin (H&E) staining was performed on the sections, which were examined using a bright-field microscope (NanoZoomer, Hamamatsu) with a 20 × 0.67 NA objective lens. All animal procedures were approved by the Institutional Animal Care and Use Committee of California Institute of Technology. We used a 2" × 2" positive 1951 USAF resolution target (58–198, Edmund Optics) to quantify the spatial resolution and DOF of our system. To prepare the thick object, carbon fibers with a diameter of 6 μm were randomly embedded in a 4% agarose block (A-204-25, GoldBio) in 3D. A 500 μm thick section was created from the agarose block using a vibratome (VT1200S, Leica). Next, the section was placed onto a standard microscope glass slide and fixed by applying cyanoacrylate glue around the edge. A coverglass was put on top of the sample and sealed using epoxy glue to prevent dehydration of the agarose.

**Data acquisition and processing**

A custom-written LabVIEW (National Instruments) program was used to synchronize the raster scanning of the 3-axis motor with the data acquisition of the time controller and acquire the raw singles and coincidence counts of the two SPCMs. When acquiring 2D imaging data, the LabVIEW program raster scanned the $x$- and $y$-axis motors and converted the raw singles counts of the signal channel and coincidence counts into classical and ICE images, respectively. The images were displayed on screen and saved to the computer in tag image file format (TIF). For imaging thick objects, multiple 2D images each captured at a $z$-position were combined to form a 3D stack. The TIF files were imported into MATLAB (MathWorks) and processed with custom-written scripts. Depending on the objects being imaged, the images were rotated, cropped, or inverted before being used to extract line profiles or edge spread functions for estimating resolution and DOF. Additionally, to compensate for the low contrast between the brain structure and the background, the brain slice images were denoised by block-matching and 3D filtering[43] followed by a variance-stabilizing transformation[44].

**Measurements of resolution and depth of field**

To measure the spatial resolution of our system, the profile of a line along $x$ perpendicular to an edge in the USAF resolution target (e.g., the yellow dashed line in Fig. 2a) was extracted and fitted to an edge spread function (ESF) centered at $x_0$, i.e., $\mathrm{ESF}(x) = a\,\mathrm{erf}((x - x_0)/w) + b$, where $a$ and $b$ are coefficients and $w$ is the radius of the beam. A Gaussian line spread function (LSF) was obtained by taking the derivative of the ESF, i.e., $\mathrm{LSF}(x) = d\mathrm{ESF}(x)/dx = 2a\exp(-(x - x_0)^2/w^2)/(w\sqrt{\pi})$. The resolution was estimated to be the FWHM of the LSF, i.e., $\mathcal{R} = 2\sqrt{\ln 2}\,w$. The mean value of the resolution was estimated to be $2\sqrt{\ln 2}$ times the fitted $w$, and the standard error was calculated to be $\sqrt{\ln 2}/1.96$ times the 95% confidence interval of the fitted $w$. To measure the DOF of our system, resolution, $\mathcal{R}$, was estimated at each $z$ position (e.g., Fig. 2d). The curves were fitted for $z_R$ to a hyperbolic function, i.e., $\mathcal{R}(z) = \mathcal{R}_0\sqrt{1 + (z - z_0)^2/z_R^2}$, where $\mathcal{R}_0$ is the focal resolution and $z_R$ is the Rayleigh length. The mean DOF was estimated to be $2z_R$, and the standard error was estimated to be 1/1.96 times the 95% confidence interval of the fitted $z_R$.

### Imaging with stray light

A white LED (MNWHL4, Thorlabs) powered by an LED driver (DC2200, Thorlabs) was used to randomly generate stray light during imaging, as shown in Supplementary Fig. 10. The LED driver was externally triggered by an analog output device (PCI-6711, National Instruments) installed on the computer. While raster scanning the object prepared on the microscope slide, at each pixel, the LabVIEW program generated a random number uniformly distributed between 0 and 1 to determine whether to trigger the white LED to output stray light. If the random number was less than 0.2, the LED was triggered to generate stray light; otherwise, no stray light would be generated. Therefore, approximately 20% of the pixels would be disrupted by stray light. To evaluate how robust the classical imaging and ICE were against stray light, we acquired images under different stray light optical powers. We calculated the structural similarity index measure (SSIM) between each image and the ground truth at zero stray light by $\mathrm{SSIM} = 4\mu_1\mu_2\sigma_{12}/\left((\mu_1^2 + \mu_2^2)(\sigma_1^2 + \sigma_2^2)\right)$, where $\mu_i$ and $\sigma_i^2$ ($i = 1$ or 2) are the average and variance of each image, respectively, and $\sigma_{12}$ is the covariance of the two images[35].

## Data availability

All data used in this study are available from the corresponding author upon reasonable request.

## Code availability

All custom codes used in this study are available from the corresponding author upon reasonable request.

## Acknowledgments

We thank Dr. David Prober and Tasha Cammidge for preparing the zebrafish specimen. We thank Dr. Lei Li for preparing the brain slice. We thank Dr. Peng Wang and Dr. Li Lin for assistance with the experiment. We also thank Dr. Kelvin Titimbo Chaparro and Siddik Suleyman Kahraman for discussion. This project has been made possible in part by grant number 2020-225832 from the Chan Zuckerberg Initiative DAF, an advised fund of Silicon Valley Community Foundation, and National Institutes of Health grants R35 CA220436 (Outstanding Investigator Award) and R01 EB028277.

## Author contributions

Y.Z., Z.H., and X.T. built the imaging system, performed the experiments, and analyzed the data. Y.Z. developed the data acquisition program. Z.H. developed the quantum imaging theory. X.T. developed the sub-shot-noise algorithms. Y.Z. and X.T. developed the ghost birefringence imaging theory and algorithms. Y.Z., Z.H., X.T., and D.C.G. prepared the manuscript. R.C. prepared the agarose-embedded zebrafish and carbon fibers. L.V.W. conceived the concept and supervised the project. All authors contributed to writing the manuscript.

## Competing interests

The authors declare no competing interests.

# Supplementary Information

# Quantum imaging of biological organisms through spatial and polarization entanglement

Yide Zhang[†], Zhe He[†], Xin Tong[†], David C. Garrett, Rui Cao, and Lihong V. Wang[*]

*Caltech Optical Imaging Laboratory, Andrew and Peggy Cherng Department of Medical Engineering, Department of Electrical Engineering, California Institute of Technology, 1200 E. California Blvd., MC 138-78, Pasadena, CA 91125, USA*

† These authors contributed equally.

* Correspondence should be addressed to L.V.W. (LVW@caltech.edu).

**Supplementary Note 1 Sub-shot-noise signal retrieval in ICE**

Each round of ICE acquisition generates three images: the signal image $N_s(\boldsymbol{r})$, the idler image $N_i(\boldsymbol{r})$, and the coincidence image $N_c(\boldsymbol{r})$. $N_s(\boldsymbol{r})$ and $N_i(\boldsymbol{r})$ contain photon counts from both SPDC photon pairs (whose averaged value is denoted as $\mu_{\text{SPDC}}$) and stray light (whose averaged value is denoted as $\mu_{\text{stray}}$). For simplicity, we assume the signal and idler detectors have the same background light intensity and detection efficiency, denoted as $\eta$.

The imaging of an object here measures its transmittance $T(\boldsymbol{r})$. Although the following derivation applies to both $N_s$ and $N_c$, we use $N_s$ as an example. Classically, the transmittance is estimated as

$$T_0(\boldsymbol{r}) = \frac{N_s(\boldsymbol{r})}{N_s^0(\boldsymbol{r})}, \tag{S1}$$

where $N_s^0$ denotes the signal image when the object is absent. In ICE, we estimate $N_s^0$ using $\langle N_s^{\text{b}}(\boldsymbol{r})\rangle_{\boldsymbol{r}}$, where $N_s^{\text{b}}$ denotes a background region of the $N_s$ image outside the target, and $\langle\ldots\rangle_{\boldsymbol{r}}$ denotes averaging over spatial locations.

By using the correlation between the SPDC photon pairs, two types of sub-shot-noise (SSN) algorithms have been adopted to enhance the SNR of the transmittance measurements. The first type relies on the ratio of the two images, where the object's transmittance is estimated as[27,45]

$$T_1(\boldsymbol{r}) = \frac{N_s(\boldsymbol{r})}{N_i(\boldsymbol{r})} \cdot \frac{\langle N_i(\boldsymbol{r})\rangle_{\boldsymbol{r}}}{\langle N_s^{\text{b}}(\boldsymbol{r})\rangle_{\boldsymbol{r}}}. \tag{S2}$$

The second type of SSN algorithms, termed optimized subtraction, suppresses the noise in $N_s$ by subtracting the variation of $N_i$[15,46]:

$$N_s^{\text{SSN}}(\boldsymbol{r}) = N_s(\boldsymbol{r}) - k(\boldsymbol{r})\Delta N_i(\boldsymbol{r}), \tag{S3}$$

where $k(\boldsymbol{r})$ is the unknown spatially varying multiplier, and $\Delta N_i(\boldsymbol{r}) = N_i(\boldsymbol{r}) - \langle N_i(\boldsymbol{r})\rangle_{\boldsymbol{r}}$. The ideal $k(\boldsymbol{r})$ is proportional to the transmittance $T(\boldsymbol{r})$, the ground truth of which is unknown. To estimate the $k(\boldsymbol{r})$, one may use the approximated transmittance $\hat{T}(\boldsymbol{r})$. For example, $\hat{T}(\boldsymbol{r})$ can be acquired using $T_0$ as in Eq. (S1) or as in Ref. [15]. The object's transmittance estimated using the second type of SSN algorithms is given by

$$T_2(\boldsymbol{r}) = \frac{N_s(\boldsymbol{r})}{\langle N_s^{\mathrm{b}}(\boldsymbol{r})\rangle_{\boldsymbol{r}}} - \hat{T}(\boldsymbol{r})\eta\left(\frac{\mu_{\mathrm{SPDC}}}{\mu_{\mathrm{stray}} + \mu_{\mathrm{SPDC}}}\right)^2 \frac{\Delta N_i(\boldsymbol{r})}{\langle N_s^{\mathrm{b}}(\boldsymbol{r})\rangle_{\boldsymbol{r}}}. \quad \text{(S4)}$$

Both Eqs. (S2) and (S4) achieve higher SNR than Eq. (S1), demonstrating the quantum advantage. However, these methods require either prior knowledge of $T(\boldsymbol{r})$ or assumptions on the photon distribution and minimal stray light intensity. Here, inspired by the two algorithms, we introduce the covariance-over-variance (CoV) algorithm to further improve the SSN performance with fewer assumptions.

The workflow of the CoV algorithm is shown in Supplementary Fig. 1. We acquire the time-lapsed image stack of $N_s(\boldsymbol{r}, t)$, $N_i(\boldsymbol{r}, t)$, and $N_c(\boldsymbol{r}, t)$. Following the basic framework of the optimized subtraction (i.e., Eq. (S3)), instead of estimating $k(\boldsymbol{r})$ with approximated $T(\boldsymbol{r})$, we derive the optimal $k(\boldsymbol{r})$ by minimizing the variance of $N_s^{\mathrm{SSN}}(\boldsymbol{r}, t)$:

$$\mathrm{Var}_t[N_s^{\mathrm{SSN}}(\boldsymbol{r}, t)] = \mathrm{Var}_t[N_s(\boldsymbol{r}, t)] + k^2(\boldsymbol{r})\mathrm{Var}_t[N_i(\boldsymbol{r}, t)] - 2k(\boldsymbol{r})\mathrm{Cov}_t[N_s(\boldsymbol{r}, t), N_i(\boldsymbol{r}, t)], \quad \text{(S5)}$$

where $\mathrm{Var}_t$ and $\mathrm{Cov}_t$ denote the variance and covariance along the time sequence, respectively. To minimize $\mathrm{Var}_t[N_s^{\mathrm{SSN}}(\boldsymbol{r}, t)]$ with regard to $k(\boldsymbol{r})$:

$$\frac{\partial \mathrm{Var}_t[N_s^{\mathrm{SSN}}(\boldsymbol{r}, t)]}{\partial k(\boldsymbol{r})} = 2k(\boldsymbol{r})\mathrm{Var}_t[N_i(\boldsymbol{r}, t)] - 2\mathrm{Cov}_t[N_s(\boldsymbol{r}, t), N_i(\boldsymbol{r}, t)] = 0. \quad \text{(S6)}$$

The optimized $k^*(\boldsymbol{r})$ is thus given by

$$k^*(\boldsymbol{r}) = \frac{\mathrm{Cov}_t[N_s(\boldsymbol{r}, t), N_i(\boldsymbol{r}, t)]}{\mathrm{Var}_t[N_i(\boldsymbol{r}, t)]}. \quad \text{(S7)}$$

Since $\langle N_s^{\mathrm{SSN}}(\boldsymbol{r})\rangle_{\boldsymbol{r}} = \langle N_s(\boldsymbol{r})\rangle_{\boldsymbol{r}}$ according to Eq. (S3), combining Eqs. (S1), (S3), and (S7) completes the CoV algorithm:

$$T_3(\boldsymbol{r}) = \frac{N_s(\boldsymbol{r})}{\langle N_s^{\mathrm{b}}(\boldsymbol{r})\rangle_{\boldsymbol{r}}} - \frac{\mathrm{Cov}_t[N_s(\boldsymbol{r}, t), N_i(\boldsymbol{r}, t)]}{\mathrm{Var}_t[N_i(\boldsymbol{r}, t)]} \cdot \frac{\Delta N_i(\boldsymbol{r})}{\langle N_s^{\mathrm{b}}(\boldsymbol{r})\rangle_{\boldsymbol{r}}}. \quad \text{(S8)}$$

From Eqs. (S5) and (S7), we can derive the minimized variance as

$$\mathrm{Var}_t[N_s^{\mathrm{SSN}}(\boldsymbol{r}, t)] = \mathrm{Var}_t[N_s(\boldsymbol{r}, t)] - \frac{\mathrm{Cov}_t^2[N_s(\boldsymbol{r}, t), N_i(\boldsymbol{r}, t)]}{\mathrm{Var}_t[N_i(\boldsymbol{r}, t)]} = \mathrm{Var}_t[N_s(\boldsymbol{r}, t)]\left(1 - \rho_{N_s, N_i}^2\right), \quad \text{(S9)}$$

where $\rho_{N_s,N_i}$ is the Pearson's correlation coefficient between $N_s$ and $N_i$ along the time sequence. Note that the variance of the classical algorithm given by Eq. (S1) is $\mathrm{Var}_t[N_s(\boldsymbol{r},t)]$. Since $\rho^2_{N_s,N_i} \geq 0$, the CoV algorithm guarantees SNR enhancement.

It is worth noting that the CoV algorithm requires repeated measurements over time. With a single-frame acquisition, we propose a similar algorithm to estimate $k(\boldsymbol{r})$ based on spatial repetitions, named as the s-CoV algorithm.

The workflow of the s-CoV algorithm is shown in Supplementary Fig. 2. From the single-frame image $N_s(\boldsymbol{r})$ (or $N_c(\boldsymbol{r})$), we calculate the histogram and divide the pixel values into $L$ bins. The selection of $L$ depends on the experimental configuration and can be optimized through iteration. For the $l$-th bin ($l = 1,2,\ldots,L$), we select the pixels from the image whose values fall into this bin and form the image subset $N_s^l(\boldsymbol{r})$. The binary mask $M^l(\boldsymbol{r})$ used for segmentation (i.e., $N_s^l(\boldsymbol{r}) = M^l(\boldsymbol{r}) \cdot N_s(\boldsymbol{r})$) is then applied to $N_i(\boldsymbol{r})$ to get the image subset $N_i^l(\boldsymbol{r})$. Following Eq. (S7), we can estimate the subset of $k^*(\boldsymbol{r})$ (denoted as $k^{n,*}$) as

$$k^{l,*} = \frac{\mathrm{Cov}_{\boldsymbol{r}}\left[N_s^l(\boldsymbol{r}), N_i^l(\boldsymbol{r})\right]}{\mathrm{Var}_{\boldsymbol{r}}[N_i^l(\boldsymbol{r})]}, \tag{S10}$$

where $\mathrm{Var}_{\boldsymbol{r}}$ and $\mathrm{Cov}_{\boldsymbol{r}}$ denote the variance and covariance along spatial locations, respectively. The same procedure is repeated for all $l$, and the resulting $k^*(\boldsymbol{r})$ is the summation of all $M^l(\boldsymbol{r})$ modified by $k^{l,*}$:

$$k^*(\boldsymbol{r}) = \sum_l k^{l,*} M^l(\boldsymbol{r}) = \sum_l \frac{\mathrm{Cov}_{\boldsymbol{r}}\left[N_s^l(\boldsymbol{r}), N_i^l(\boldsymbol{r})\right]}{\mathrm{Var}_{\boldsymbol{r}}[N_i^l(\boldsymbol{r})]} M^l(\boldsymbol{r})\,. \tag{S11}$$

Combining Eqs. (S1), (S3), and (S11) completes the s-CoV algorithm:

$$T_4(\boldsymbol{r}) = \frac{N_s(\boldsymbol{r})}{\langle N_s^{\mathrm{b}}(\boldsymbol{r})\rangle_{\boldsymbol{r}}} - \left[\sum_l \frac{\mathrm{Cov}_{\boldsymbol{r}}\left[N_s^l(\boldsymbol{r}), N_i^l(\boldsymbol{r})\right]}{\mathrm{Var}_{\boldsymbol{r}}[N_i^l(\boldsymbol{r})]} M^l(\boldsymbol{r})\right] \cdot \frac{\Delta N_i(\boldsymbol{r})}{\langle N_s^{\mathrm{b}}(\boldsymbol{r})\rangle_{\boldsymbol{r}}}. \tag{S12}$$

To compare the performances of the three algorithms, we simulate the 1D case where the object is placed in the signal arm with constant $T(x) = 0.5$. Each detector in the signal and idler arms performs Bernoulli trials with probability $\eta$ to select the SPDC photons (following a Poisson distribution with $\mu_{\mathrm{SPDC}}$). The stray light photons in each detector follow two independent Poisson distributions with $\eta\mu_{\mathrm{stray}}$. The schematics are shown in Supplementary Fig. 3a.

We simulate two scenarios with (1) a fixed stray light-SPDC light ratio ($\mu_{\mathrm{stray}}/\mu_{\mathrm{SPDC}} = 1$) and a varying detector efficiency $\eta$ and (2) a fixed $\eta = 0.7$ and a varying $\mu_{\mathrm{stray}}/\mu_{\mathrm{SPDC}}$. Supplementary Figs. 3b and c show the SNR enhancement using $N_s$ and $N_i$ images with Eqs. (S2), (S4), and (S8) compared to Eq. (S1). Supplementary Figs. 3d and e show similar results using $N_c$ and $N_i$ images. In all cases, the CoV algorithm outperforms the others consistently.

**Supplementary Note 2 Entanglement pinhole in ICE**

In the simplified schematic of the imaging system shown in Supplementary Fig. 5, $\boldsymbol{r}_{0,s}$, $\boldsymbol{r}_{1,s}$, and $\boldsymbol{r}_{2,s}$ represent the coordinates of the BBO, the object, and the two detectors $D_s$ for signal photons, respectively. $\boldsymbol{r}_{0,i}$ and $\boldsymbol{r}_{2,i}$ represent the transverse coordinates of the BBO and the detector $D_i$ for idler photons, respectively. $\boldsymbol{k}_{0,s}$ and $\boldsymbol{k}_{0,i}$ represent the wavevectors of the entangled signal and idler photons emitted from the BBO, respectively. $\boldsymbol{k}_{1,s}$ denotes the wavevector of the signal photon after the first objective. $\boldsymbol{k}'_{1,s}$ denotes the wavevector of the signal photon emitted from the object. $\boldsymbol{k}_{2,s}$ denotes the wavevector of the signal photon on the detector $D_s$. All wavevectors have the same constant magnitude $k$. Here, the subscripts *s* and *i* denote signal and idler, respectively.

The rate of coincidence counts of the two detectors is given by

$$G_{\mathrm{ICE}}^{(2)}(\boldsymbol{r}_{2,s}, \boldsymbol{r}_{2,i}) = \left|\left\langle 0\left|\hat{E}_s^{(+)}\hat{E}_i^{(+)}\right|\xi\right\rangle\right|^2, \tag{S13}$$

where $|\xi\rangle$ is the wavefunction of entangled photon pairs from the BBO:

$$|\xi\rangle = \sum_{\boldsymbol{k}_{0,s}} e^{-j\boldsymbol{k}_{0,s}\cdot\boldsymbol{r}_{0,s}} e^{-j\boldsymbol{k}_{0,i}\cdot\boldsymbol{r}_{0,i}} \left|1_{\boldsymbol{k}_{0,s}}, 1_{\boldsymbol{k}_{0,i}}\right\rangle. \tag{S14}$$

The phase-matching condition constrains that the transverse components of $\boldsymbol{k}_{0,s}$ and $\boldsymbol{k}_{0,i}$ are opposite while their axial components are identical. If $\boldsymbol{r_p}$ and $\boldsymbol{k_p}$ represent the position and wavevector of the pump light, the state denotes a spatially entangled state with $(\boldsymbol{r}_{0,s} + \boldsymbol{r}_{0,i})/2 = \boldsymbol{r_p}$ and $\boldsymbol{k}_{0,s} + \boldsymbol{k}_{0,i} = \boldsymbol{k_p}$. Because of the latter constraint, we choose to sum over $\boldsymbol{k}_{0,s}$ only. In the signal arm, $\hat{O}^s$ and $\hat{O}^{s'}$ denote the operators for the objective and lenses before and after the object plane, respectively. In the idler arm, $\hat{O}^i$ denotes the operators for the lens in front of $D_i$. For simplicity, we set the image magnification ratios of both channels to unity. $\hat{E}_s^{(-)}$ and $\hat{E}_i^{(-)}$ are the Hermitian conjugates of the electric fields $\hat{E}_s^{(+)}$ and $\hat{E}_i^{(+)}$, respectively. The electric fields are derived by propagation from the source to the detectors as follows if the amplitude transmission coefficient of the object $t = 1$ [47]:

$$\hat{E}_s^{(+)}(\boldsymbol{r}_{2,s}, \boldsymbol{r}_{1,s}, \boldsymbol{r}_{0,s}) = E_0(\boldsymbol{r}_{0,s})\hat{e}\left|0_{\boldsymbol{r}_{2,s}}\right\rangle\left\langle 1_{\boldsymbol{r}_{2,s}}\right|\hat{O}^{s'}\left|1_{\boldsymbol{r}_{1,s}}\right\rangle\left\langle 1_{\boldsymbol{r}_{1,s}}\right|\hat{O}^s. \tag{S15}$$

Then we add $t(\boldsymbol{r}_{1,s}, \boldsymbol{k}_{1,s})$ and expand Eq. (S15) over $\boldsymbol{k}_{0,s}, \boldsymbol{k}_{1,s}, \boldsymbol{k}'_{1,s}$, and $\boldsymbol{k}_{2,s}$:

$$\hat{E}_s^{(+)}(\boldsymbol{r}_{2,s},\boldsymbol{r}_{1,s},\boldsymbol{r}_{0,s}) = \sum_{\boldsymbol{k}_{0,s},\boldsymbol{k}_{1,s},\boldsymbol{k}'_{1,s},\boldsymbol{k}_{2,s}} E_0(\boldsymbol{r}_{0,s})\hat{e}|0_{\boldsymbol{r}_{2,s}}\rangle\langle 1_{\boldsymbol{r}_{2,s}}||1_{\boldsymbol{k}_{2,s}}\rangle\langle 1_{\boldsymbol{k}_{2,s}}|\hat{O}^{s'}|1_{\boldsymbol{k}'_{1,s}}\rangle$$
$$\times \langle 1_{\boldsymbol{k}'_{1,s}}||1_{\boldsymbol{r}_{1,s}}\rangle t(\boldsymbol{r}_{1,s},\boldsymbol{k}_{1,s})\langle 1_{\boldsymbol{r}_{1,s}}||1_{\boldsymbol{k}_{1,s}}\rangle\langle 1_{\boldsymbol{k}_{1,s}}|\hat{O}^{s}|1_{\boldsymbol{k}_{0,s}}\rangle\langle 1_{\boldsymbol{k}_{0,s}}|$$
$$= \sum_{\boldsymbol{k}_{0,s},\boldsymbol{k}_{1,s},\boldsymbol{k}'_{1,s},\boldsymbol{k}_{2,s}} E_0(\boldsymbol{r}_{0,s})\hat{e}|0_{\boldsymbol{r}_{2,s}}\rangle\langle 1_{\boldsymbol{r}_{2,s}}||1_{\boldsymbol{k}_{2,s}}\rangle\hat{O}^{s'}_{\boldsymbol{k}_{2,s},\boldsymbol{k}'_{1,s}}$$
$$\times \langle 1_{\boldsymbol{k}'_{1,s}}||1_{\boldsymbol{r}_{1,s}}\rangle t(\boldsymbol{r}_{1,s},\boldsymbol{k}_{1,s})\langle 1_{\boldsymbol{r}_{1,s}}||1_{\boldsymbol{k}_{1,s}}\rangle\hat{O}^{s}_{\boldsymbol{k}_{1,s},\boldsymbol{k}_{0,s}}\langle 1_{\boldsymbol{k}_{0,s}}|$$
$$= \sum_{\boldsymbol{k}_{0,s},\boldsymbol{k}_{1,s},\boldsymbol{k}'_{1,s},\boldsymbol{k}_{2,s}} E_0(\boldsymbol{r}_{0,s})\hat{e}t(\boldsymbol{r}_{1,s},\boldsymbol{k}_{1,s})|0_{\boldsymbol{r}_{2,s}}\rangle h(\boldsymbol{r}_{2,s},\boldsymbol{k}_{2,s};\boldsymbol{r}_{1,s},\boldsymbol{k}'_{1,s})$$
$$\times \langle 1_{\boldsymbol{r}_{1,s}}||1_{\boldsymbol{k}_{1,s}}\rangle\hat{O}^{s}_{\boldsymbol{k}_{1,s},\boldsymbol{k}_{0,s}}\langle 1_{\boldsymbol{k}_{0,s}}|, \tag{S16}$$

$$\hat{E}_i^{(+)}(\boldsymbol{r}_{2,i}) = E_0\hat{e}|0_{\boldsymbol{r}_{2,i}}\rangle\langle 1_{\boldsymbol{r}_{2,i}}|\hat{O}^i = \sum_{\boldsymbol{k}_{0,i},\boldsymbol{k}_{2,i}} E_0\hat{e}|0_{\boldsymbol{r}_{2,i}}\rangle\langle 1_{\boldsymbol{r}_{2,i}}||1_{\boldsymbol{k}_{2,i}}\rangle\langle 1_{\boldsymbol{k}_{2,i}}|\hat{O}^i|1_{\boldsymbol{k}_{0,i}}\rangle\langle 1_{\boldsymbol{k}_{0,i}}|$$
$$= \sum_{\boldsymbol{k}_{0,i},\boldsymbol{k}_{2,i}} E_0\hat{e}|0_{\boldsymbol{r}_{2,i}}\rangle\langle 1_{\boldsymbol{r}_{2,i}}||1_{\boldsymbol{k}_{2,i}}\rangle\hat{O}^i_{\boldsymbol{k}_{2,i},\boldsymbol{k}_{0,i}}\langle 1_{\boldsymbol{k}_{0,i}}|. \tag{S17}$$

Here, $E_0$ is the amplitude of the electric field; $\hat{e}$ is the polarization unit vector. The projectors $|0_{\boldsymbol{r}_{2,s}}\rangle\langle 1_{\boldsymbol{r}_{2,s}}|$ and $|0_{\boldsymbol{r}_{2,i}}\rangle\langle 1_{\boldsymbol{r}_{2,i}}|$ account for measurements at $\boldsymbol{r}_{2,s}$ and $\boldsymbol{r}_{2,i}$, respectively. The difference between $\boldsymbol{k}_{1,s}$ and $\boldsymbol{k}'_{1,s}$ accounts for the diffraction of the object. The type-I SPDC crystal guarantees the same polarization for the signal and idler photons. $h(\boldsymbol{r}_{2,s},\boldsymbol{k}_{2,s};\boldsymbol{r}_{1,s},\boldsymbol{k}'_{1,s}) = \langle 1_{\boldsymbol{r}_{2,s}}||1_{\boldsymbol{k}_{2,s}}\rangle\hat{O}^{s'}_{\boldsymbol{k}_{2,s},\boldsymbol{k}'_{1,s}}\langle 1_{\boldsymbol{k}'_{1,s}}||1_{\boldsymbol{r}_{1,s}}\rangle = e^{i\phi^s_{12}}$, and the phase shift $\phi^s_{12}$ is related to the signal photon propagation from the object to the detector $D_s$. The phase changes $\phi^s_{01}$ and $\phi^i_{02}$ are given in a similar way: $h(\boldsymbol{r}_{1,s},\boldsymbol{k}_{1,s};\boldsymbol{r}_{0,s},\boldsymbol{k}_{0,s}) = \langle 1_{\boldsymbol{r}_{1,s}}||1_{\boldsymbol{k}_{1,s}}\rangle\hat{O}^{s}_{\boldsymbol{k}_{1,s},\boldsymbol{k}_{0,s}}\langle 1_{\boldsymbol{k}_{0,s}}||1_{\boldsymbol{r}_{0,s}}\rangle = e^{i\phi^s_{01}}$, and $h(\boldsymbol{r}_{2,i},\boldsymbol{k}_{2,i};\boldsymbol{r}_{0,i},\boldsymbol{k}_{0,i}) = \langle 1_{\boldsymbol{r}_{2,i}}||1_{\boldsymbol{k}_{2,i}}\rangle\hat{O}^{i}_{\boldsymbol{k}_{2,i},\boldsymbol{k}_{0,i}}\langle 1_{\boldsymbol{k}_{0,i}}||1_{\boldsymbol{r}_{0,s}}\rangle = e^{i\phi^i_{02}}$.

Next, we substitute $\hat{E}_s^{(+)}$, $\hat{E}_i^{(+)}$, and $|\xi\rangle$ into Eq. (S13):

$$\hat{E}_i^{(+)}(\boldsymbol{r}_{2,i},\boldsymbol{r}_{0,i})\hat{E}_s^{(+)}(\boldsymbol{r}_{2,s},\boldsymbol{r}_{1,s},\boldsymbol{r}_{0,s})|\xi\rangle$$
$$= E_0^2 \sum_{\boldsymbol{k}_{0,s},\boldsymbol{k}_{1,s},\boldsymbol{k}'_{1,s},\boldsymbol{k}_{2,s}} t(\boldsymbol{r}_{1,s},\boldsymbol{k}_{1,s})|0_{\boldsymbol{r}_{2,s}},0_{\boldsymbol{r}_{2,i}}\rangle$$
$$\times h(\boldsymbol{r}_{2,s},\boldsymbol{k}_{2,s};\boldsymbol{r}_{1,s},\boldsymbol{k}'_{1,s})h(\boldsymbol{r}_{1,s},\boldsymbol{k}_{1,s};\boldsymbol{r}_{0,s},\boldsymbol{k}_{0,s})h(\boldsymbol{r}_{2,i},\boldsymbol{k}_{2,i};\boldsymbol{r}_{0,i},\boldsymbol{k}_{0,i}). \tag{S18}$$

Next, to simplify Eq. (S18), we define

$$t'(\boldsymbol{r}_{1,s}) = \frac{\sum_{\boldsymbol{k}_{0,s},\boldsymbol{k}_{1,s}} t(\boldsymbol{r}_{1,s},\boldsymbol{k}_{1,s}) h(\boldsymbol{r}_{1,s},\boldsymbol{k}_{1,s};\boldsymbol{r}_{0,s},\boldsymbol{k}_{0,s}) h(\boldsymbol{r}_{2,i},\boldsymbol{k}_{2,i};\boldsymbol{r}_{0,i},\boldsymbol{k}_{0,i})}{\sum_{\boldsymbol{k}_{0,s},\boldsymbol{k}_{1,s}} h(\boldsymbol{r}_{1,s},\boldsymbol{k}_{1,s};\boldsymbol{r}_{0,s},\boldsymbol{k}_{0,s}) h(\boldsymbol{r}_{2,i},\boldsymbol{k}_{0,i};\boldsymbol{r}_{0,i},\boldsymbol{k}_{0,i})}. \quad \text{(S19)}$$

Substitute Eqs. (S18) and (S19) into Eq. (S13):

$$G_{\mathrm{ICE}}^{(2)}(\boldsymbol{r}_{0,i},\boldsymbol{r}_{1,s};\boldsymbol{r}_{2,s},\boldsymbol{r}_{2,i}) = E_0^4 |t'(\boldsymbol{r}_{1,s})|^2 \left| \sum_{\boldsymbol{k}'_{1,s},\boldsymbol{k}_{2,s}} h(\boldsymbol{r}_{2,s},\boldsymbol{k}_{2,s};\boldsymbol{r}_{1,s},\boldsymbol{k}'_{1,s}) \right|^2$$

$$\times \left| \sum_{\boldsymbol{k}_{0,s},\boldsymbol{k}_{1,s}} h(\boldsymbol{r}_{1,s},\boldsymbol{k}_{1,s};\boldsymbol{r}_{0,s},\boldsymbol{k}_{0,s}) h(\boldsymbol{r}_{2,i},\boldsymbol{k}_{2,i};\boldsymbol{r}_{0,i},\boldsymbol{k}_{0,i}) \right|^2. \quad \text{(S20)}$$

The detectors detect true coincidence of a photon pair within the coincidence window (8 ns). Photon pairs from different positions of the BBO most likely fall into different coincidence windows and, hence, are considered incoherent[48]. Integrating the contributions of photon pairs from different positions incoherently yields the total rate of coincidence counts:

$$G_{\mathrm{ICE}}^{(2)}(\boldsymbol{r}_{1,s};\boldsymbol{r}_{2,s},\boldsymbol{r}_{2,i}) = \int_S p_0(\boldsymbol{r}_{0,s}) G_{\mathrm{ICE}}^{(2)}(\boldsymbol{r}_{0,i},\boldsymbol{r}_{1,s};\boldsymbol{r}_{2,s},\boldsymbol{r}_{2,i}) d\boldsymbol{r}_{0,s}$$

$$= E_0^4 |t'(\boldsymbol{r}_{1,s})|^2 \left| \sum_{\boldsymbol{k}'_{1,s},\boldsymbol{k}_{2,s}} h(\boldsymbol{r}_{2,s},\boldsymbol{k}_{2,s};\boldsymbol{r}_{1,s},\boldsymbol{k}'_{1,s}) \right|^2$$

$$\times \int_S p_0(\boldsymbol{r}_{0,s}) \left| \sum_{\boldsymbol{k}_{0,s},\boldsymbol{k}_{1,s}} h(\boldsymbol{r}_{1,s},\boldsymbol{k}_{1,s};\boldsymbol{r}_{0,s},\boldsymbol{k}_{0,s}) h(\boldsymbol{r}_{2,i},\boldsymbol{k}_{2,i};\boldsymbol{r}_{0,i},\boldsymbol{k}_{0,i}) \right|^2 d\boldsymbol{r}_{0,s}. \quad \text{(S21)}$$

Here, $S$ represents the spatial domain of the BBO source, and $p_0(\boldsymbol{r}_{0,s})$ denotes the probability density function for each photon pair at the source. To further simplify the result, we define the following PSFs:

$$h'_{\mathrm{ep}} = \sum_{\boldsymbol{k}_{0,s},\boldsymbol{k}_{1,s}} h(\boldsymbol{r}_{1,s},\boldsymbol{k}_{1,s};\boldsymbol{r}_{0,s},\boldsymbol{k}_{0,s}) h(\boldsymbol{r}_{2,i},\boldsymbol{k}_{2,i};\boldsymbol{r}_{0,i},\boldsymbol{k}_{0,i})$$

$$= \sum_{\boldsymbol{k}_{0,s},\boldsymbol{k}_{1,s}} e^{i\phi_{01}^s} e^{i\phi^i} = \sum_{\boldsymbol{k}_{0,s},\boldsymbol{k}_{1,s}} e^{i\phi_{\mathrm{ep}}}, \quad \text{(S22)}$$

where $\phi_{\mathrm{ep}} = \phi_{01}^s + \phi^i$ is the equivalent phase shift related to $h'_{\mathrm{ep}}$.

$$|h_{\mathrm{ep}}(\boldsymbol{r}_{2,i};\boldsymbol{r}_{1,s})|^2 = \int_S p_0(\boldsymbol{r}_{0,s}) |h'_{\mathrm{ep}}|^2 d\boldsymbol{r}_{0,s}, \quad \text{(S23)}$$

$$h_s(\boldsymbol{r}_{1,s};\boldsymbol{r}_{2,s}) = \sum_{\boldsymbol{k}'_{1,s},\boldsymbol{k}_{2,s}} h(\boldsymbol{r}_{2,s},\boldsymbol{k}_{2,s};\boldsymbol{r}_{1,s},\boldsymbol{k}'_{1,s}), \quad \text{(S24)}$$

where the subscript ep denotes the entanglement pinhole. Consequently, we reach

$$G^{(2)}_{\mathrm{ICE}}(\boldsymbol{r}_{1,s};\boldsymbol{r}_{2,s},\boldsymbol{r}_{2,i}) = E_0^4\left|t'(\boldsymbol{r}_{1,s})\right|^2\left|h_s(\boldsymbol{r}_{1,s};\boldsymbol{r}_{2,s})\right|^2\left|h_{\mathrm{ep}}(\boldsymbol{r}_{2,i};\boldsymbol{r}_{1,s})\right|^2. \quad \text{(S25)}$$

Integration over the finite apertures of the two detectors yields

$$G^{(2)}_{\mathrm{ICE}}(\boldsymbol{r}_{1,s}) = E_0^4\left|t'(\boldsymbol{r}_{1,s})\right|^2 \int_{D_s} p_s(\boldsymbol{r}_{2,s})\left|h_s(\boldsymbol{r}_{1,s};\boldsymbol{r}_{2,s})\right|^2 d\boldsymbol{r}_{2,s}$$
$$\times \int_{D_i} p_i(\boldsymbol{r}_{2,i})\left|h_{\mathrm{ep}}(\boldsymbol{r}_{2,i};\boldsymbol{r}_{1,s})\right|^2 d\boldsymbol{r}_{2,i}, \quad \text{(S26)}$$

where $p_s(\boldsymbol{r}_{2,s})$ and $p_i(\boldsymbol{r}_{2,i})$ denote the photon detection probability density functions for the two detectors, respectively.

For classical imaging (CI) using signal-channel-only detection (i.e., raw singles photon counts received by $D_s$), the photon counting rate is

$$G^{(1)}_{\mathrm{CI}}(\boldsymbol{r}_{2,s}) = \left|\left\langle 0\left|\hat{E}_s^{(+)}\right|\xi_s\right\rangle\right|^2, \quad \text{(S27)}$$

where the signal photon state $|\xi_s\rangle = \sum_{\boldsymbol{k}_{0,s}} e^{-j\boldsymbol{k}_{0,s}\cdot\boldsymbol{r}_{0,s}}\left|1_{\boldsymbol{k}_{0,s}}\right\rangle$. With $\hat{E}_s^{(+)}$and $|\xi_s\rangle$ we find

$$\hat{E}_s^{(+)}(\boldsymbol{r}_{2,s},\boldsymbol{r}_{1,s})|\xi_s\rangle = E_0$$
$$\times \sum_{\boldsymbol{k}_{0,s},\boldsymbol{k}_{1,s},\boldsymbol{k}'_{1,s},\boldsymbol{k}_{2,s}} \left|0_{\boldsymbol{r}_{2,s}}\right\rangle t(\boldsymbol{r}_{1,s},\boldsymbol{k}_{1,s}) h_s(\boldsymbol{r}_{2,s},\boldsymbol{k}_{2,s};\boldsymbol{r}_{1,s},\boldsymbol{k}'_{1,s}) h_s(\boldsymbol{r}_{1,s},\boldsymbol{k}_{1,s};\boldsymbol{r}_{0,s},\boldsymbol{k}_{0,s}). \quad \text{(S28)}$$

Hence, we derive and simplify $G^{(1)}_{\mathrm{CI}}(\boldsymbol{r}_{1,s};\boldsymbol{r}_{2,s})$ to be

$$G^{(1)}_{\mathrm{CI}}(\boldsymbol{r}_{1,s};\boldsymbol{r}_{2,s}) = E_0^2\left|t'(\boldsymbol{r}_{1,s})\right|^2\left|h_s(\boldsymbol{r}_{1,s};\boldsymbol{r}_{2,s})\right|^2 \times \int_S p_0(\boldsymbol{r}_{0,s})\left|A_{s0}h_s(\boldsymbol{r}_{0,s};\boldsymbol{r}_{1,s})\right|^2 d\boldsymbol{r}_{0,s}. \quad \text{(S29)}$$

Integration over the finite aperture of the detector yields

$$G^{(1)}_{\mathrm{CI}}(\boldsymbol{r}_{1,s};\boldsymbol{r}_{2,s})$$
$$= E_0^2\left|t'(\boldsymbol{r}_{1,s})\right|^2 \int_{D_s} p_s(\boldsymbol{r}_{2,s})\left|h_s(\boldsymbol{r}_{1,s};\boldsymbol{r}_{2,s})\right|^2 d\boldsymbol{r}_{2,s} \times \int_S p_0(\boldsymbol{r}_{0,s})\left|h_s(\boldsymbol{r}_{0,s};\boldsymbol{r}_{1,s})\right|^2 d\boldsymbol{r}_{0,s}. \quad \text{(S30)}$$

If we assume that the object in the experiment is not sensitive to the incidence wavevectors such that $t(\boldsymbol{r}_{1,s},\boldsymbol{k}_{1,s}) \approx t(\boldsymbol{r}_{1,s},\boldsymbol{k}_0)$, and $\boldsymbol{k}_0$ is along the optical axis, Eq. (S19) indicates $t'(\boldsymbol{r}_{1,s}) = t(\boldsymbol{r}_{1,s})$.

**Supplementary Note 3 ICE using accidental coincidences**

If the two detectors only measure accidental coincidences, the results could be treated as coincidences of photons from different positions $(\boldsymbol{r}_{0,s}, \boldsymbol{r}_{0,i})$ with different wavevectors $(\boldsymbol{k}_{0,s}, \boldsymbol{k}_{0,i})$. The source state is replaced by

$$|\xi\rangle_{\text{acc}} = \sum_{\boldsymbol{k}_{0,s},\boldsymbol{k}_{0,i}} e^{-j\boldsymbol{k}_{0,s}\cdot\boldsymbol{r}_{0,s}} e^{-j\boldsymbol{k}_{0,i}\cdot\boldsymbol{r}_{0,i}} \left|1_{\boldsymbol{k}_{0,s}}, 1_{\boldsymbol{k}_{0,i}}\right\rangle, \tag{S31}$$

which can be written as a product state $|\xi\rangle_{\text{acc}} = \sum_{\boldsymbol{k}_{0,s}} e^{-j\boldsymbol{k}_{0,s}\cdot\boldsymbol{r}_{0,s}} \left|1_{\boldsymbol{k}_{0,s}}\right\rangle \otimes \sum_{\boldsymbol{k}_{0,i}} e^{-j\boldsymbol{k}_{0,i}\cdot\boldsymbol{r}_{0,i}} \left|1_{\boldsymbol{k}_{0,i}}\right\rangle$. Eq. (S21) becomes

$$\begin{aligned} &G^{(2)}_{\text{acc}}(\boldsymbol{r}_{1,s}; \boldsymbol{r}_{2,s}, \boldsymbol{r}_{2,i}) \\ &= E_0^4 \left|t'(\boldsymbol{r}_{1,s})\right|^2 \left|\sum_{\boldsymbol{k}'_{1,s},\boldsymbol{k}_{2,s}} h(\boldsymbol{r}_{2,s}, \boldsymbol{k}_{2,s}; \boldsymbol{r}_{1,s}, \boldsymbol{k}'_{1,s})\right|^2 \\ &\times \int_S p_0(\boldsymbol{r}_{0,s}) \left|\sum_{\boldsymbol{k}_{0,s},\boldsymbol{k}_{1,s}} h(\boldsymbol{r}_{1,s}, \boldsymbol{k}_{1,s}; \boldsymbol{r}_{0,s}, \boldsymbol{k}_{0,s})\right|^2 d\boldsymbol{r}_{0,s} \\ &\times \int_S p_0(\boldsymbol{r}_{0,i}) \left|\sum_{\boldsymbol{k}_{0,i},\boldsymbol{k}_{2,i}} h(\boldsymbol{r}_{2,i}, \boldsymbol{k}_{2,i}; \boldsymbol{r}_{0,i}, \boldsymbol{k}_{0,i})\right|^2 d\boldsymbol{r}_{0,i}\,. \end{aligned} \tag{S32}$$

Next, we can use $h(\boldsymbol{r}_{2,s}; \boldsymbol{r}_{1,s}) = \sum_{\boldsymbol{k}'_{1,s},\boldsymbol{k}_{2,s}} h(\boldsymbol{r}_{2,s}, \boldsymbol{k}_{2,s}; \boldsymbol{r}_{1,s}, \boldsymbol{k}'_{1,s})$, $h(\boldsymbol{r}_{1,s}; \boldsymbol{r}_{0,s}) = \sum_{\boldsymbol{k}_{0,s},\boldsymbol{k}_{1,s}} h(\boldsymbol{r}_{1,s}, \boldsymbol{k}_{1,s}; \boldsymbol{r}_{0,s}, \boldsymbol{k}_{0,s})$, and $h(\boldsymbol{r}_{2,i}; \boldsymbol{r}_{0,i}) = \sum_{\boldsymbol{k}_{0,i}} h(\boldsymbol{r}_{2,i}, \boldsymbol{k}_{2,i}; \boldsymbol{r}_{0,i}, \boldsymbol{k}_{0,i})$ to simplify $G^{(2)}_{\text{acc}}$:

$$\begin{aligned} G^{(2)}_{\text{acc}}(\boldsymbol{r}_{1,s}; \boldsymbol{r}_{2,s}, \boldsymbol{r}_{2,i}) = E_0^4 \left|t'(\boldsymbol{r}_{1,s})\right|^2 \left|h(\boldsymbol{r}_{2,s}; \boldsymbol{r}_{1,s})\right|^2 \int_S p_0(\boldsymbol{r}_{0,s}) \left|h(\boldsymbol{r}_{1,s}; \boldsymbol{r}_{0,s})\right|^2 d\boldsymbol{r}_{0,s} \\ \times \int_S p_0(\boldsymbol{r}_{0,i}) \left|h(\boldsymbol{r}_{2,i}; \boldsymbol{r}_{0,i})\right|^2 d\boldsymbol{r}_{0,i}\,. \end{aligned} \tag{S33}$$

Integration over the finite aperture of the detector yields

$$\begin{aligned} G^{(2)}_{\text{acc}}(\boldsymbol{r}_{1,s}) = E_0^4 \left|t'(\boldsymbol{r}_{1,s})\right|^2 \int_{D_s} p_s(\boldsymbol{r}_{2,s}) \left|h(\boldsymbol{r}_{2,s}; \boldsymbol{r}_{1,s})\right|^2 d\boldsymbol{r}_{2,s} \int_S p_0(\boldsymbol{r}_{0,s}) \left|h(\boldsymbol{r}_{1,s}; \boldsymbol{r}_{0,s})\right|^2 d\boldsymbol{r}_{0,s} \\ \times \int_{D_i} \int_S p_i(\boldsymbol{r}_{2,i}) p_0(\boldsymbol{r}_{0,i}) \left|h(\boldsymbol{r}_{2,i}; \boldsymbol{r}_{0,i})\right|^2 d\boldsymbol{r}_{0,i}\, d\boldsymbol{r}_{2,i}\,. \end{aligned} \tag{S34}$$

Because last term is a constant with a determined $D_i$ and $S$, $G^{(2)}_{\text{acc}} \propto G^{(1)}_{\text{CI}}$. Imaging with accidental coincidences therefore provides the same resolution and DOF as classical imaging.

**Supplementary Note 4 Characterization of polarization entanglement through Bell's test**

Bell-type inequalities provide a standard to characterize a system's ability to generate entangled states. These inequalities are constructed in favor of the local hidden variable theory (LHVT), while the violation of them as predicted by quantum mechanics is generally observed in experiments. Among the various means of Bell's tests, the Clauser–Horne–Shimony–Holt (CHSH) inequality serves as a practical benchmark[38].

Defining the Hilbert space for Alice as $\mathcal{A}$ and for Bob as $\mathcal{B}$, we denote eigenstates for the measurement axes $\hat{\alpha}$ and $\hat{\beta}$ as $A_\alpha$ and $B_\beta$, and the corresponding eigenvalues as $a$ and $b$. An LHVT suggests a hidden variable $\lambda$ with a probability density function $p(\lambda)$. For measurement outcomes $a$ and $b$, their joint probability is

$$P_{\mathrm{LHV}}(a,b|\alpha,\beta)=\int P(a|\alpha,\lambda)P(b|\beta,\lambda)p(\lambda)d\lambda, \quad \text{(S35)}$$

where $P(a|\alpha,\lambda)$ and $P(b|\beta,\lambda)$ are probabilities for Alice to obtain $a$ and Bob to obtain $b$, respectively. Now Alice and Bob set two analyzers with random angles of $\alpha$ and $\beta$. $E(\alpha,\beta)$ represents the correlation of the measurement:

$$E(\alpha,\beta)\equiv P(H,H|\alpha,\beta)+P(V,V|\alpha,\beta)-P(H,V|\alpha,\beta)-P(V,H|\alpha,\beta). \quad \text{(S36)}$$

The Bell–CHSH inequality is then given by

$$S_{\mathrm{CHSH}}=|E(\alpha,\beta)+E(\alpha',\beta)-E(\alpha,\beta')+E(\alpha',\beta')|\leq 2, \quad \text{(S37)}$$

where $\alpha'$ and $\beta'$ denote the second choices of the analyzer angles.

In contrast, quantum theory predicts that $S_{\mathrm{CHSH}}>2$ is possible with specific combinations of observation angles. According to quantum mechanics, we can model coincidence counts $\widehat{N}(H,H|\alpha,\beta)$ for a maximally entangled Bell state (i.e., the EPR state) as[37]

$$\widehat{N}(H,H|\alpha,\beta)=N_0\cos^2(\alpha-\beta)+N_1, \quad \text{(S38)}$$

where $N_0$ is the maximum true coincidence count, and $N_1$ represents the contribution of accidental coincidences. For such states, quantum theory predicts a maximum violation of Eq. (S37) at $(0°, 22.5°, 45°, 67.5°)$ at the Tsirelson's bound, $S_{\mathrm{CHSH}}^{\max}=2\sqrt{2}$.

In our experimental setup, for each round of Bell's test, we rotated $\alpha$ from 0° to 180° with a step size of 45°. For each fixed $\alpha$, we rotated $\beta$ from 0° to 180° with a step size of 22.5°. After recording the coincidence counts $N(H, H|\alpha, \beta)$ at each step with an acquisition time of 1 s and a coincidence detection window of 8 ns, we calculated the correlation value adapted from Eq. (S36) as

$$E(\alpha, \beta) = \frac{N(H,H|\alpha,\beta)+N(H,H|\alpha+90°,\beta+90°)-N(H,H|\alpha+90°,\beta)-N(H,H|\alpha,\beta+90°)}{N(H,H|\alpha,\beta)+N(H,H|\alpha+90°,\beta+90°)+N(H,H|\alpha+90°,\beta)+N(H,H|\alpha,\beta+90°)}. \quad \text{(S39)}$$

The CHSH $S$ value was then evaluated based on the value of $E$ according to Eq. (S37). The results are shown in Supplementary Fig. 12.

By performing Bell's test, our system shows a strong violation of the CHSH inequality with $S = 2.78 \pm 0.01 > 2$ estimated by calculating the mean and standard error of $S$ values measured from 10 rounds of Bell's tests. This result, which violates the Bell-CHSH inequality by more than 57 standard errors of the mean, indicates significant deviations of our result from the LHVT prediction.

**Supplementary Note 5 Polarization entanglement-enabled ghost birefringence imaging**

The polarization entanglement of the SPDC photons in the ICE system enables ghost birefringence imaging. By preparing the EPR state as described in Eq. (S38) and recording the coincidence counts, ICE can be used to measure the transmittance $T$ and the birefringence properties $\theta$ and $\Delta$ of the object, where $\theta$ is the angle of the principal refractive index axis and $\Delta$ is the phase retardation between the two refractive index axes. Here, we use Stokes vectors $(I, Q, U, V)$ and Mueller matrices to describe the state of polarization. The birefringence properties of the object can be denoted using a Mueller matrix[49]:

$$X_{\Delta,\theta} = \begin{bmatrix} 1 & 0 & 0 & 0 \\ 0 & \cos^2 2\theta + \sin^2 2\theta \cos\Delta & \cos 2\theta \sin 2\theta (1-\cos\Delta) & -\sin 2\theta \sin\Delta \\ 0 & \cos 2\theta \sin 2\theta (1-\cos\Delta) & \sin^2 2\theta + \cos^2 2\theta \cos\Delta & \cos 2\theta \sin\Delta \\ 0 & \sin 2\theta \sin\Delta & -\cos 2\theta \sin\Delta & \cos\Delta \end{bmatrix}. \quad \text{(S40)}$$

Under the EPR state, we kept $\alpha = 0°$ in the signal arm while rotating $\beta$ in the idler arm from $0°$ to $135°$ with a step size of $45°$. The corresponding polarization states of the coincidence measurements can be represented by the Stokes vectors

$$\vec{S}_{\beta=0°} = \begin{bmatrix} I_{0°} \\ Q_{0°} \\ U_{0°} \\ V_{0°} \end{bmatrix} = X_{\Delta,\theta} \begin{bmatrix} 1 \\ 1 \\ 0 \\ 0 \end{bmatrix} T, \quad \text{(S41)}$$

$$\vec{S}_{\beta=45°} = \begin{bmatrix} I_{45°} \\ Q_{45°} \\ U_{45°} \\ V_{45°} \end{bmatrix} = X_{\Delta,\theta} \begin{bmatrix} 1 \\ 0 \\ 1 \\ 0 \end{bmatrix} T, \quad \text{(S42)}$$

$$\vec{S}_{\beta=90°} = \begin{bmatrix} I_{90°} \\ Q_{90°} \\ U_{90°} \\ V_{90°} \end{bmatrix} = X_{\Delta,\theta} \begin{bmatrix} 1 \\ -1 \\ 0 \\ 0 \end{bmatrix} T, \quad \text{(S43)}$$

$$\vec{S}_{\beta=135°} = \begin{bmatrix} I_{135°} \\ Q_{135°} \\ U_{135°} \\ V_{135°} \end{bmatrix} = X_{\Delta,\theta} \begin{bmatrix} 1 \\ 0 \\ -1 \\ 0 \end{bmatrix} T. \quad \text{(S44)}$$

Substituting Eq. (S40) into Eqs. (S41)–(S44), we have

$$I_{0°} = T, Q_{0°} = T(\cos^2 2\theta + \sin^2 2\theta \cos\Delta),$$
$$I_{45°} = T,\ Q_{45°} = T \cos 2\theta \sin 2\theta (1-\cos\Delta),$$
$$I_{90°} = T, Q_{90°} = -T(\cos^2 2\theta + \sin^2 2\theta \cos\Delta),$$
$$I_{135°} = T,\ Q_{135°} = -T\cos 2\theta \sin 2\theta (1-\cos\Delta).$$

The coincidence counts for $\beta = 0°, 45°, 90°, 135°$, therefore, can be represented as

$$N_{0°} = \frac{1}{2}(I_{0°} + Q_{0°}) = \frac{T}{2}(1 + \cos^2 2\theta + \sin^2 2\theta \cos\Delta), \tag{S45}$$

$$N_{45°} = \frac{1}{2}(I_{45°} + Q_{45°}) = \frac{T}{2}(1 + \cos 2\theta \sin 2\theta(1 - \cos\Delta)), \tag{S46}$$

$$N_{90°} = \frac{1}{2}(I_{90°} + Q_{90°}) = \frac{T}{2}(1 - \cos^2 2\theta - \sin^2 2\theta \cos\Delta), \tag{S47}$$

$$N_{135°} = \frac{1}{2}(I_{135°} + Q_{135°}) = \frac{T}{2}(1 - \cos 2\theta \sin 2\theta(1 - \cos\Delta)). \tag{S48}$$

Consequently, based on Eqs. (S45)–(S48), the transmittance and birefringence properties of the object can be extracted from the coincidence counts:

$$T = \frac{1}{2}(N_{0°} + N_{45°} + N_{90°} + N_{135°}), \tag{S49}$$

$$\theta = \frac{1}{2}\tan^{-1}\left(\frac{2N_{90°}}{N_{45°} - N_{135°}}\right), \tag{S50}$$

$$\Delta = \cos^{-1}\left(1 - \frac{(N_{45°} - N_{135°})^2 + 4N_{90°}^2}{N_{90°}(N_{0°} + N_{45°} + N_{90°} + N_{135°})}\right). \tag{S51}$$

In classical imaging, the birefringence properties of the object need to be measured using incident photons with different polarization states[50]. In ICE, however, birefringence imaging can be performed without changing the polarization states of the photons incident on the object. When the polarization of the signal photons was kept constant ($\alpha = 0°$) while the polarization states of the idler photons were varied ($\beta = 0°, 45°, 90°, 135°$), the classical images acquired with the raw signal counts showed no differences, unable to extract the birefringence properties of the object (Supplementary Fig. 13a). In comparison, the ICE images acquired with the coincidence counts exhibited substantial differences following Eqs. (S45)–(S48) (Supplementary Fig. 13b), which could be used to extract the transmittance and birefringence properties of the object (Supplementary Fig. 13c). One may regard this approach as quantum "ghost birefringence imaging". Enabled by polarization entanglement, the ghost birefringence quantification of ICE demonstrates a true quantum advantage over classical imaging.

**Supplementary Note 6 Imaging by coincidence with a classical light source.**

Classical two-photon coincidence imaging can be achieved when the spatially entangled source is replaced with a classical pulse source[39]. We implemented such imaging as shown in Supplementary Fig. 15, where a 635-nm CW laser (MLL-III-635-100mW, CNI Laser) was modulated at 4 kHz by a mechanical chopper (MC1F60, Thorlabs). The modulated beam was split by a beam splitter (BS013, Thorlabs) and sent to the signal and idler arms of the ICE system in Fig. 1, where the polarization selectors (HWP and PBS) were removed. The images formed with the raw signal counts using the quantum and classical sources are shown in Supplementary Figs. 16a and b, respectively. For a fair comparison, we used a neutral density filter to attenuate the classical beam such that it provided the same photon flux to the object as that of the SPDC signal beam (~19 kHz at maximum transmittance). Because the SPDC beam contained more spatial modes than the classical beam, the raw signal image (i.e., the classical image) generated with the quantum source exhibited a lower spatial resolution than the one generated with the classical source.

Using the quantum source, the ICE image in Supplementary Fig. 16c showed a maximum coincidence count rate at 485 Hz with an SNR of 22. In comparison, using the classical source, the ICE image in Supplementary Fig. 16d exhibited a maximum coincidence count rate at 16 Hz with an SNR of 4, which was 5.5 times lower than the SNR from the quantum source. Therefore, whereas it is possible to use classical correlation to generate ICE images, the SNR of the image is substantially lower than that generated with a spatially entangled quantum source. Consequently, to generate ICE images with the same SNR, the object needs to be illuminated with a classical source that is 30 times stronger than that of a quantum source, which could cause damage to photosensitive biological samples. Moreover, classical correlation is incompatible with either the sub-shot-noise algorithms (Supplementary Note 1), which require a spatially entangled quantum source, or the ghost birefringence quantification that is enabled by polarization entanglement (Supplementary Note 5). Therefore, the quantum correlation of hyperentangled SPDC photons in ICE is advantageous over classical correlation, especially for imaging photosensitive biological samples.

**Supplementary Note 7 Comparison of ICE, GI, and CPI.**

Here, we compare ICE (Supplementary Fig. 17a) with two existing quantum imaging methods. Quantum ghost imaging (GI) utilizes spatially entangled photons to record an image of an object using photons that have not interacted with the object[40]. A typical ghost imaging setup uses entangled SPDC photons generated by a nonlinear medium, e.g., a BBO crystal, and exploits the spatial correlations between the positions of the photon pairs in the signal and idler arms (Supplementary Fig. 17b). The signal photons interact with the object and is detected by a non-spatially resolving bucket detector. The idler photons are detected by a multi-pixel camera, which, upon coincidence detection of signal and idler photons, provides a ghost image of the object. It is noted that neither signal nor idler beams alone contain enough information to reconstruct an image of the object. However, the spatial entanglement between the signal and idler photons can be utilized to extract the image.

As an extension of GI, correlation plenoptic imaging (CPI) is another quantum imaging method that utilizes spatial correlations of photon pairs[41]. Beyond the position correlation utilized in GI, CPI also exploits the momentum correlation of the photon pairs, capturing the light field (position and direction of the light) emanating from the object, thus allowing refocusing, DOF extension, and 3D visualization. Unlike GI, which requires a bucket detector and a camera, CPI utilizes two well-aligned multi-pixel cameras to simultaneously record the position and momentum of the photon pairs. Thus far, CPI has only been demonstrated experimentally with chaotic light[51]. Although achieving CPI with entangled SPDC photons has been proved theoretically[52], it has not been demonstrated experimentally. A theoretical framework of CPI with entangled photons adopts a multi-pixel camera in the idler arm to record the positions of the object, which, upon coincidence detection with the camera in the signal arm, provides a ghost image of the object (Supplementary Fig. 17c). Additionally, through a lens that conjugates the BBO crystal and the camera in the signal arm, the momentum of the photons at each pixel of the object is recorded. The simultaneous recording of the position and momentum of the spatially entangled photons enables the reconstruction of a plenoptic image of the object.

Despite the similarity in using entangled photon pairs and coincidence detection for imaging, ICE is fundamentally different from GI or CPI (Supplementary Fig. 17). First, ICE directly images the object by focusing the SPDC beam onto the object and recording the coincidence of two single-pixel detectors (SPCMs) while raster scanning the object. The raster scanning extends the resolvable pixel counts indefinitely, enabling quantum imaging over a large FOV. In comparison, both GI and CPI provide indirect ghost images of the object by triggering a multi-pixel camera using either a bucket detector or another camera. Because the multi-pixel cameras have a limited number of resolvable pixel counts, the FOVs of GI and CPI are limited. Second, ICE provides spatial resolution of the object, so it is capable of classically imaging the object using the signal arm alone. GI and CPI, however, do not provide spatial resolution of the object, and hence cannot image the object by only using the signal arm. Third, owing to the focusing and single-pixel detection of the SPDC beam, ICE measures substantially more spatial modes per pixel than GI and CPI, where the modes of the SPDC beam are evenly distributed across the multi-pixel cameras. The larger number of spatial modes per pixel leads to a higher SNR of ICE over GI or CPI under the same photon flux. Fourth, whereas ICE, GI, and CPI all utilize only spatially entangled photon pairs generated from BBO crystals, ICE also exploits the polarization entanglement of the photon pairs for ghost birefringence imaging of the object (Supplementary Note 5).

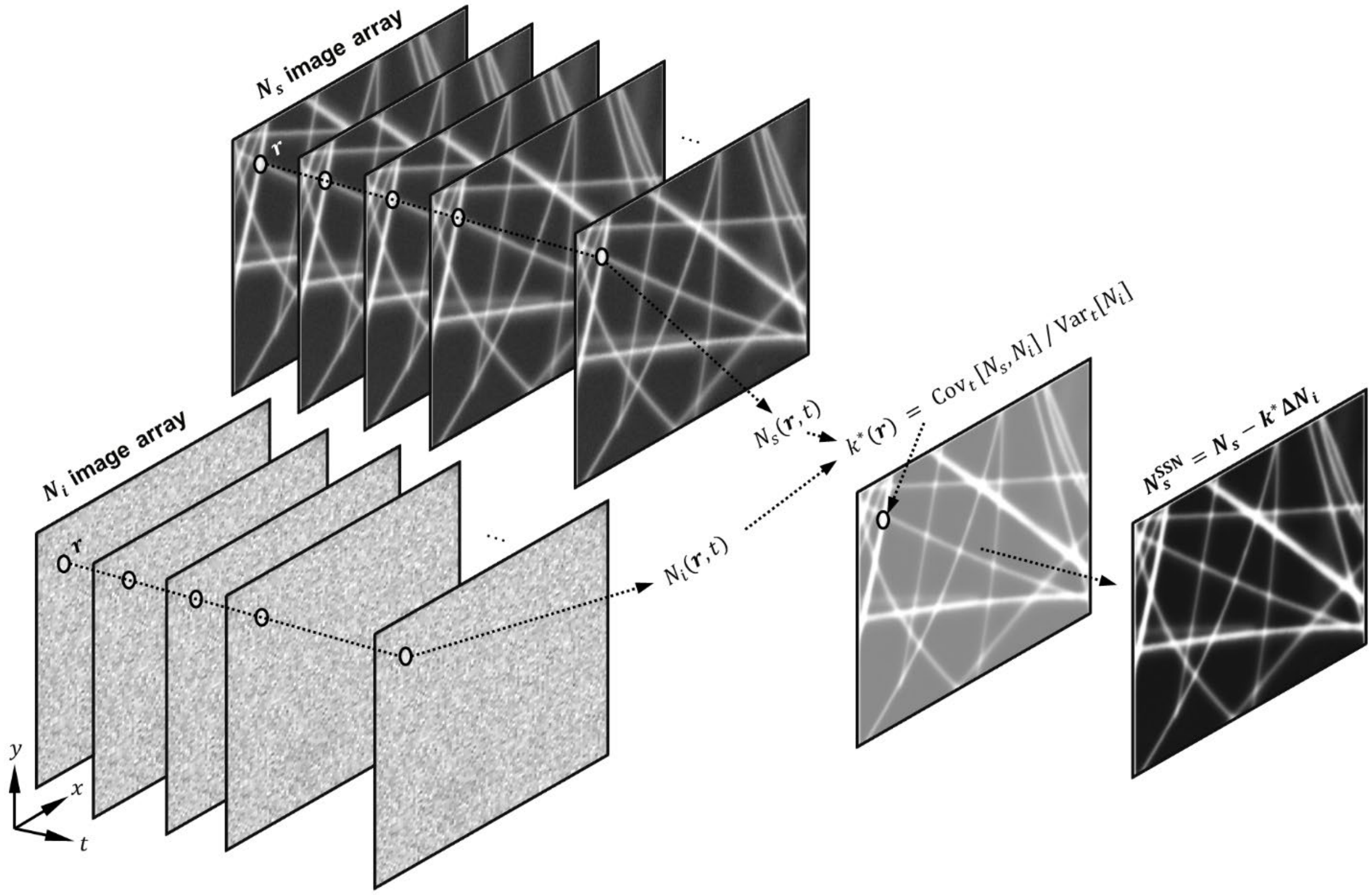

**Supplementary Fig. 1 Illustration of the CoV algorithm.**

Each pixel $\boldsymbol{r}$ from the image stacks $N_s(\boldsymbol{r},t)$ and $N_i(\boldsymbol{r},t)$ form two time sequences, and their temporal covariance and variance are computed to generate $k^*(\boldsymbol{r})$ using Eq. (S7). The $k^*(\boldsymbol{r})$ image is used in the CoV algorithm to form the sub-shot-noise image $N_s^{\mathrm{SSN}}(\boldsymbol{r})$ using Eq. (S8).

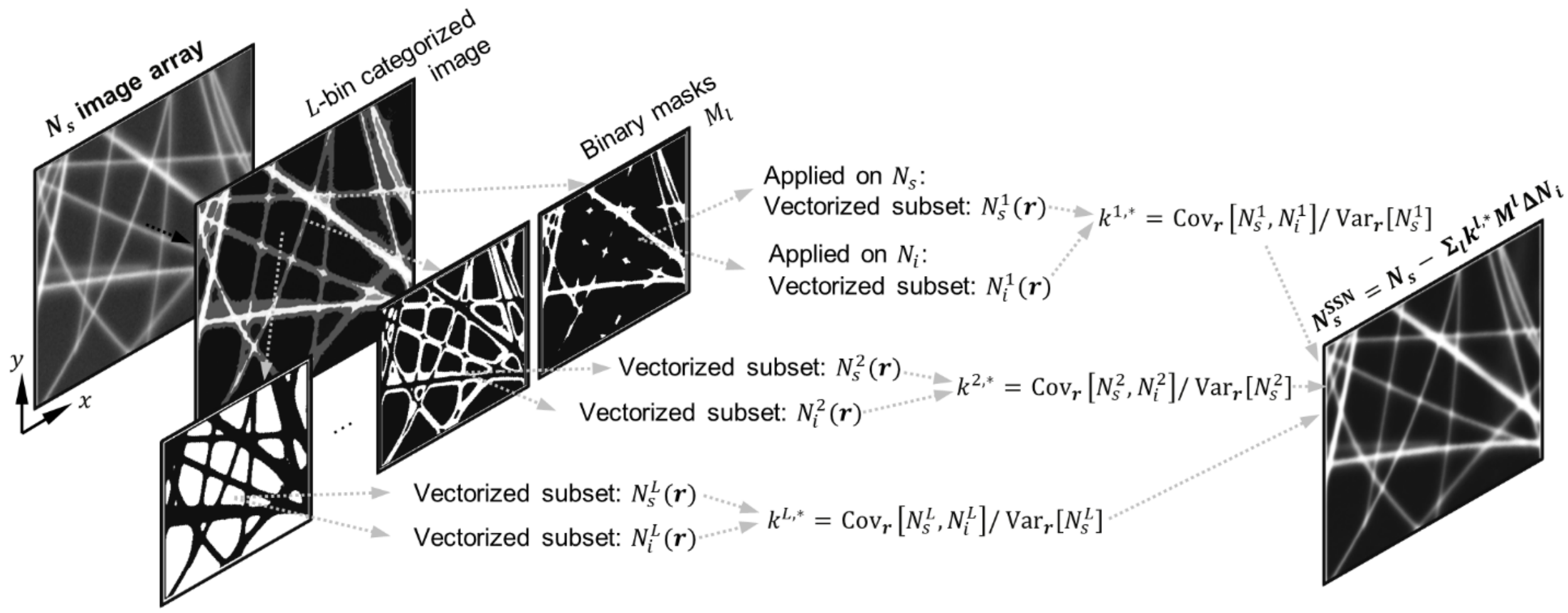

**Supplementary Fig. 2 Illustration of the s-CoV algorithm.**

The images $N_s(\boldsymbol{r})$ and $N_i(\boldsymbol{r})$ are divided into $L$ subset image pairs according to the pixel values in the $N_s$ image. For each image pair $N_s^l$ and $N_i^l$, the $k^{l,*}$ value is computed through Eq. (S10). The same procedure is repeated for all image pairs to form the $k^*(\boldsymbol{r})$ image, which is used in the s-CoV algorithm to form the sub-shot-noise image $N_s^{\mathrm{SSN}}(\boldsymbol{r})$ using Eq. (S12).

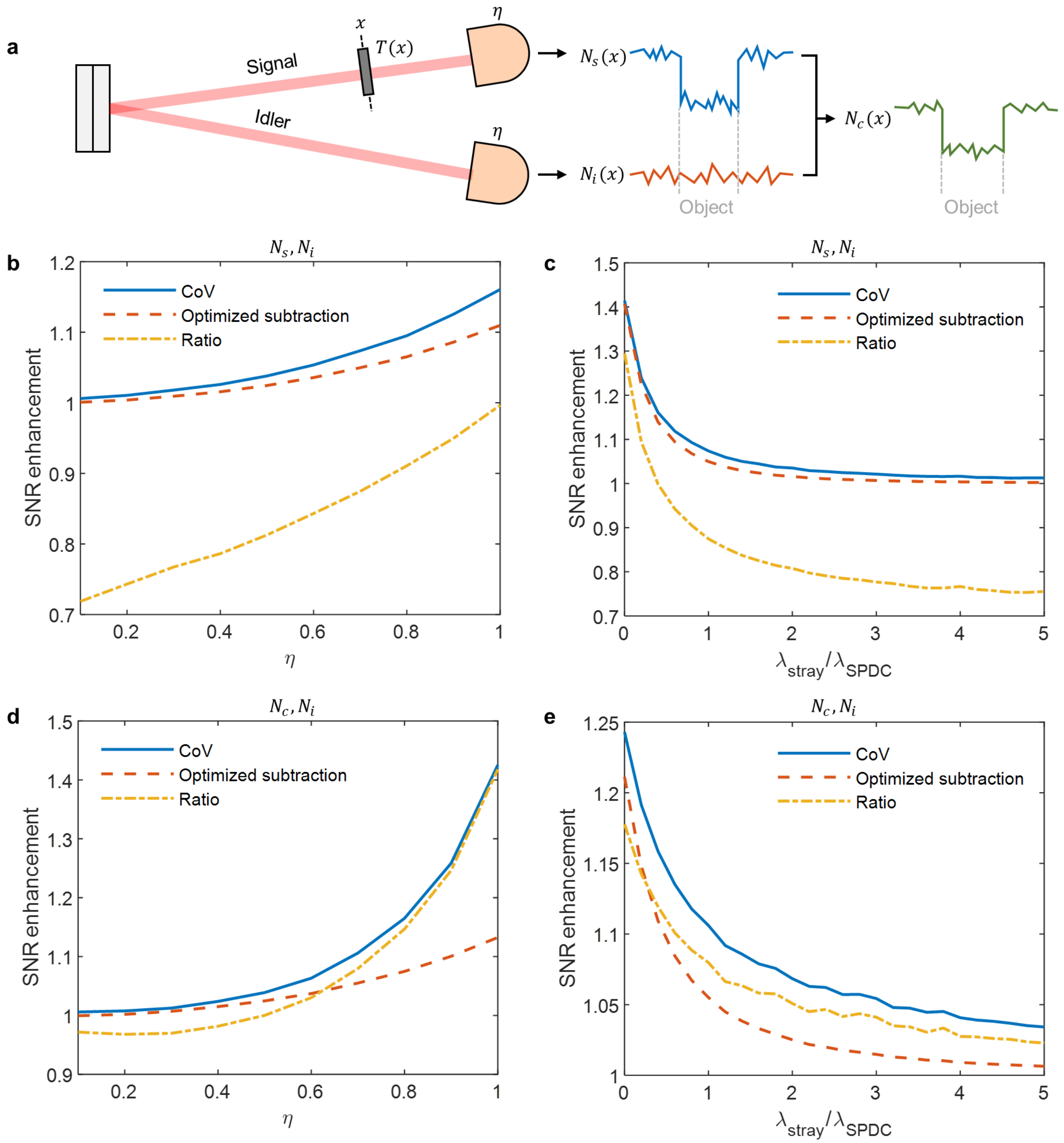

**Supplementary Fig. 3 SSN signal retrieval simulations.**

**a,** Schematics of the simulated setup. 1D object with transmittance $T(x) = 0.5$ is placed in the signal arm. $N_s(x)$ and $N_i(x)$ denote the images measured by detectors in the signal and idler arms. $\eta$ is the detection efficiency. The $N_c(x)$ image is extracted from $N_s$ and $N_i$. **b,** SNR enhancement over classical measurement (Eq. (S1)) simulated using the ratio (Eq. (S2)), optimized subtraction (Eq. (S4)), and CoV (Eq. (S8)) algorithms with $N_s$ and $N_i$. The ratio of stray light and SPDC photon counts is fixed to 1, and the detector efficiency $\eta$ is varied. **c,** Simulated SNR enhancement from $N_s$ and $N_i$ with the detector efficiency $\eta$ fixed at 0.7 and the ratio of stray light and SPDC photon counts varied. **d,** Simulated SNR enhancement from $N_c$ and $N_i$ with the ratio of stray light and SPDC photon counts fixed to 1 and the detector efficiency $\eta$ varied. **e,** Simulated SNR enhancement from $N_c$ and $N_i$ with the detector efficiency $\eta$ fixed at 0.7 and the ratio of stray light and SPDC photon counts varied.

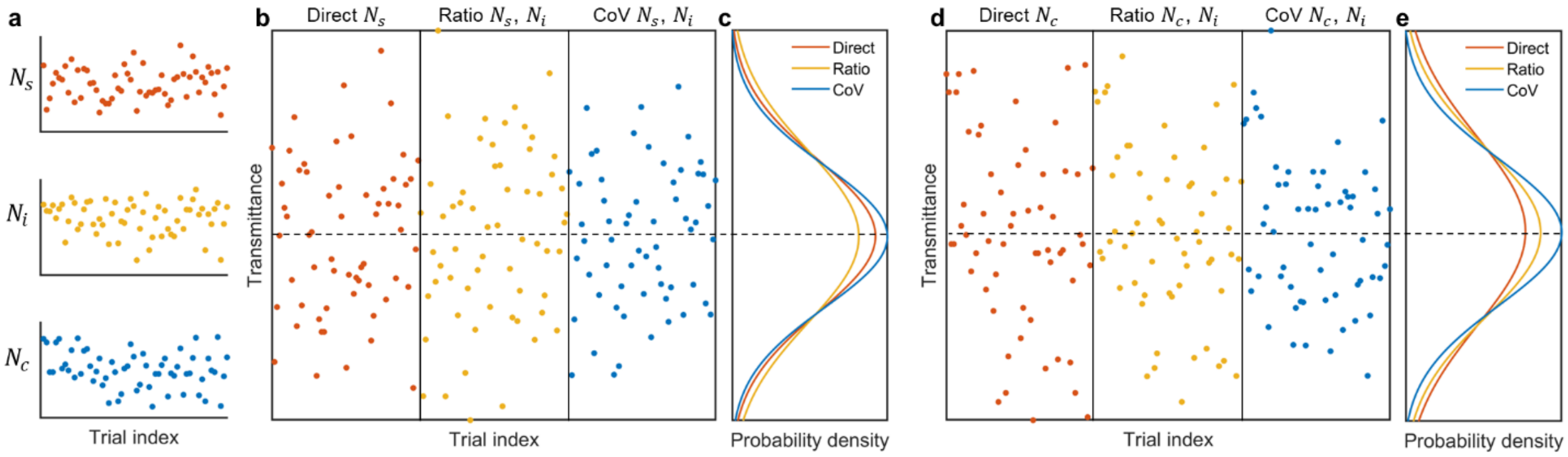

**Supplementary Fig. 4 SSN signal retrieval experiments.**

**a**, Signal $N_s$, idler $N_i$, and coincidence $N_c$ counts acquired from a series of trials. **b**, Transmittance of the object measured using $N_s$ directly, the ratio of $N_s$ to $N_i$, and the CoV algorithm on $N_s$ and $N_i$. **c**, Histograms of the transmittance measured in **b**. **d**, Transmittance of the object measured using $N_c$ directly, the ratio of $N_c$ to $N_i$, and the CoV algorithm on $N_c$ and $N_i$. **e**, Histograms of the transmittance measured in **d**.

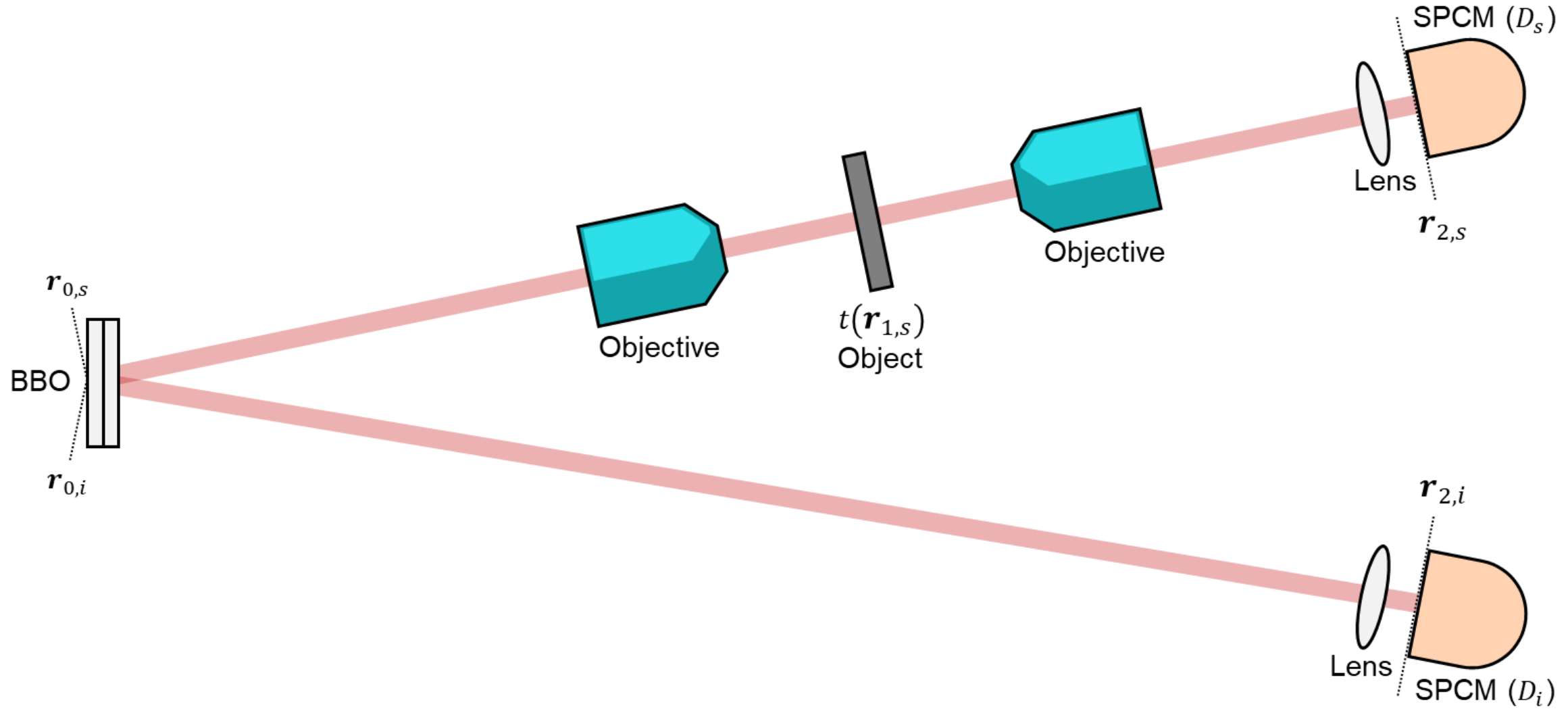

**Supplementary Fig. 5 Illustration of the ICE model.**

SPCM, single photon counting module. $\boldsymbol{r}_{0,s}$, signal photon position on the source. $\boldsymbol{r}_{0,i}$, idler photon position on the source. $D_s$, $D_i$, detectors. $t(\boldsymbol{r}_{1,s})$, amplitude transmission coefficient of the object. $\boldsymbol{r}_{1,s}$, the position on the object. $\boldsymbol{r}_{2,s}$ and $\boldsymbol{r}_{2,i}$, the positions on the detectors $D_s$ and $D_i$, respectively.

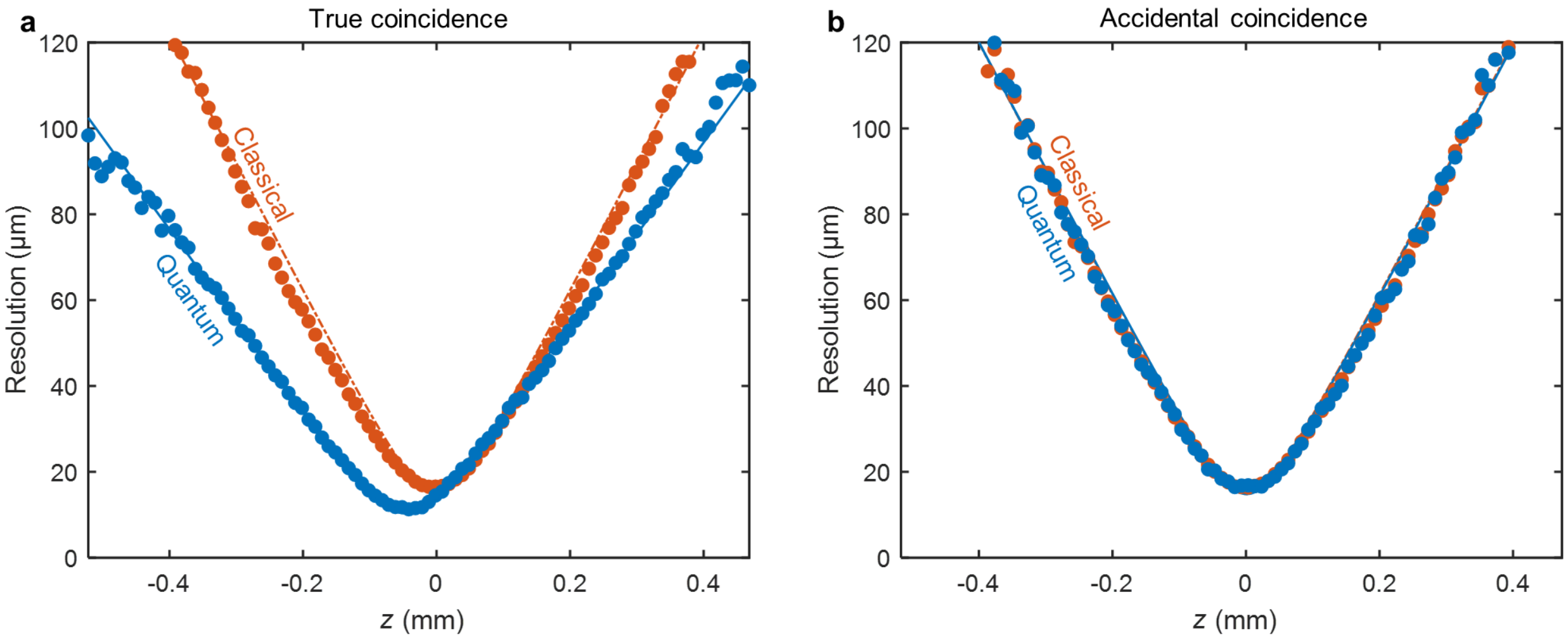

**Supplementary Fig. 6 ICE with true and accidental coincidences.**

Resolution versus $z$ measured with true coincidences (**a**) and accidental coincidences (**b**). Dots represent experimental measurements. Solid and dash-dotted lines denote fits. The ICE measurements require the coincidence window to be 8 ns. For the accidental ICE measurements, the coincidence window was set to 400 ns to allow accidental coincidences. The results are explained in Supplementary Note 3.

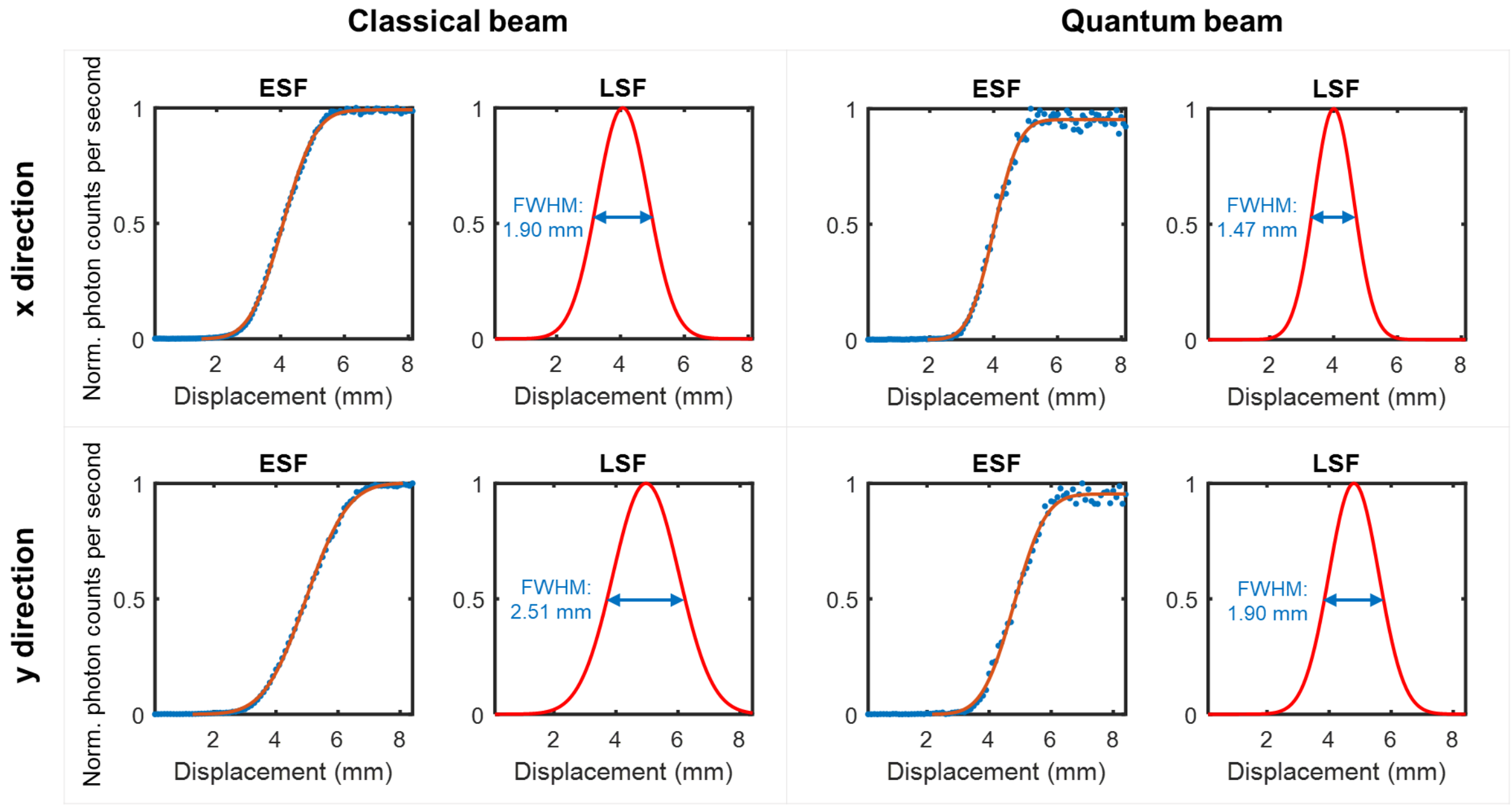

**Supplementary Fig. 7 Characterization of horizontal ($x$) and vertical ($y$) beam widths of classical imaging and ICE.**

ESFs were acquired by scanning a sharp edge to block the beams. The FWHMs of the LSFs (derivatives of the ESFs) were used to estimate the beam widths.

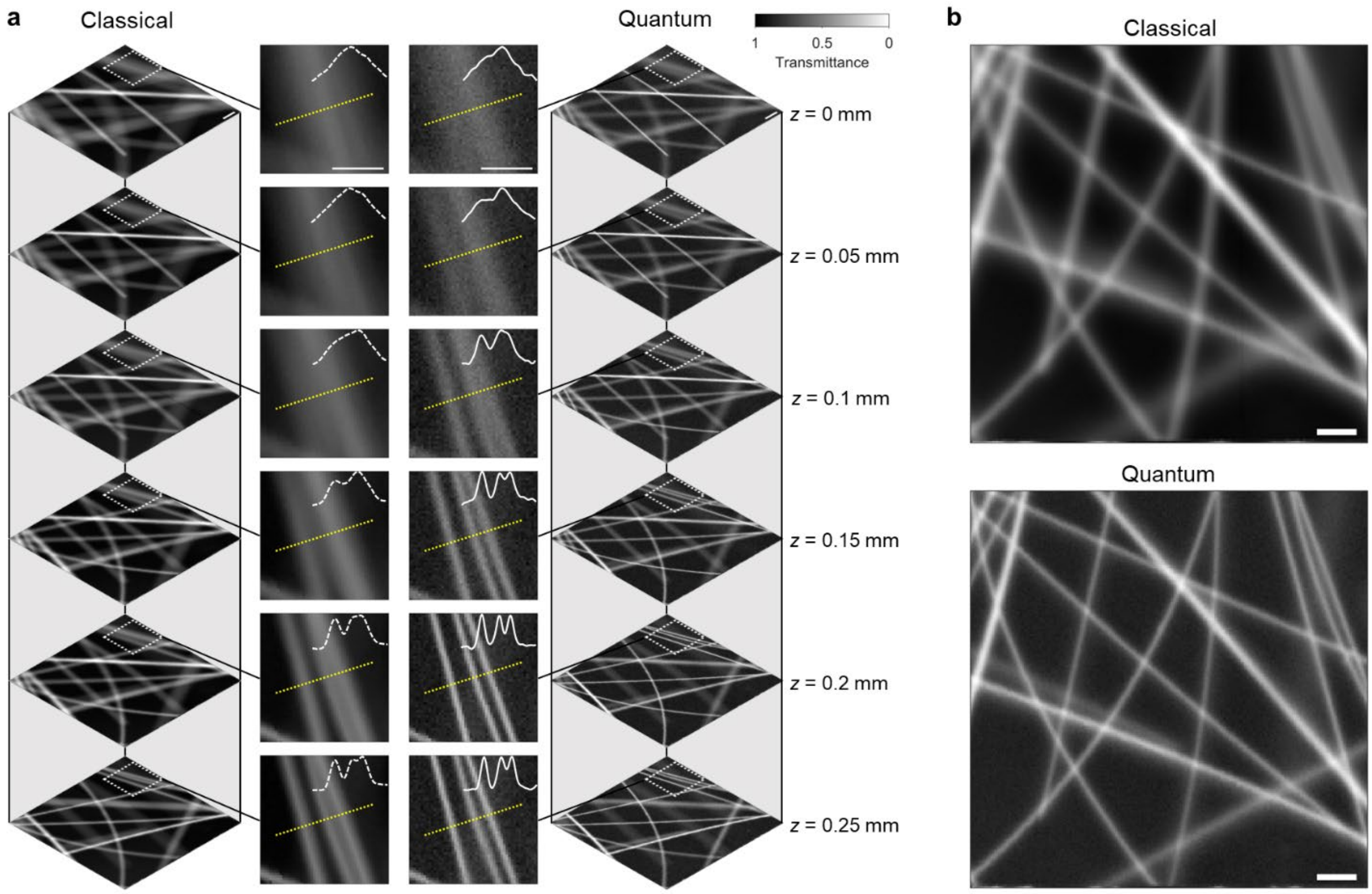

**Supplementary Fig. 8 ICE of carbon fibers embedded in thick agarose.**

**a**, Classical and ICE images of carbon fibers embedded in agarose at different *z* positions. Profiles of the yellow dotted lines are plotted in the closeups. **b**, Average of the stacks in **a**. Norm., normalized. Scale bars, 100 μm.

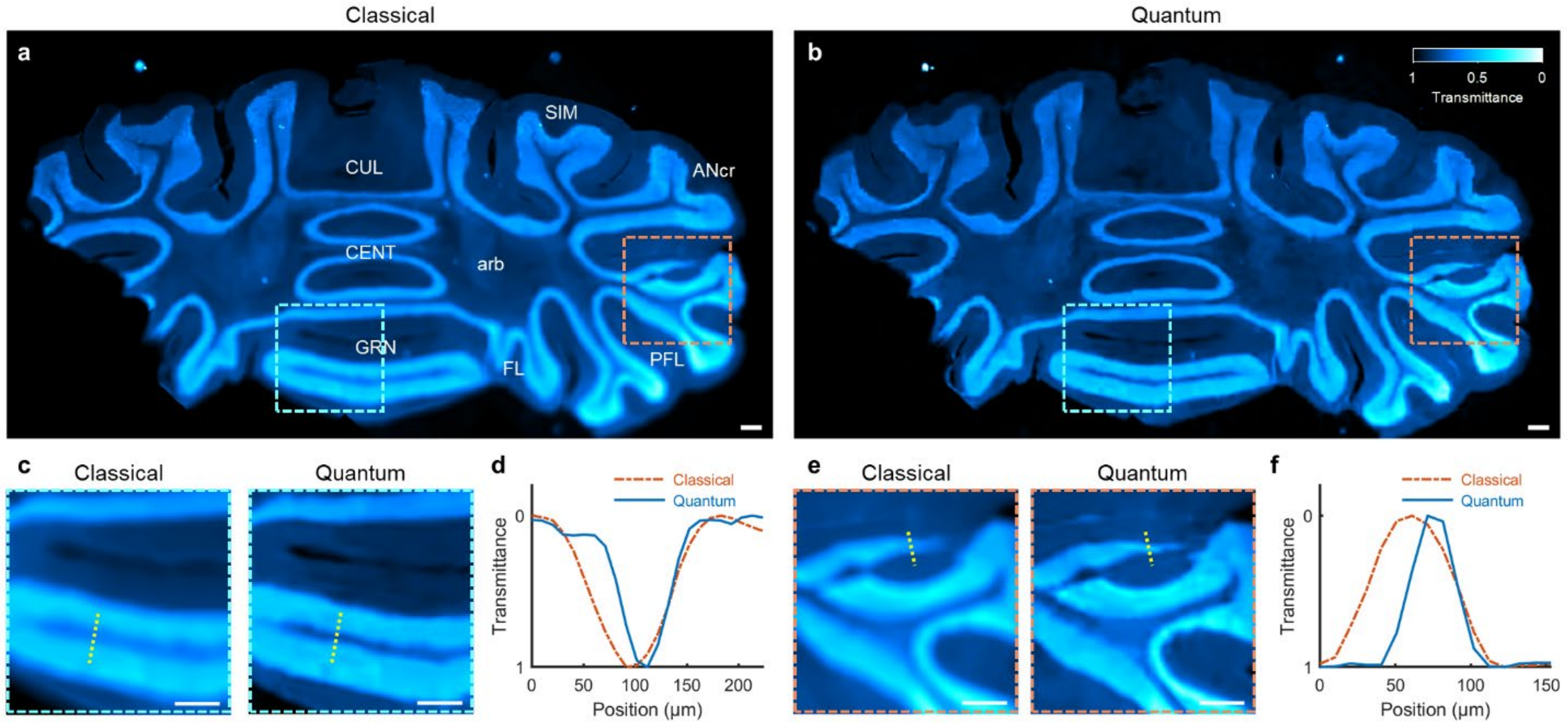

**Supplementary Fig. 9 ICE of a mouse brain slice.**

**a**,**b**, Classical (**a**) and ICE (**b**) images of a hematoxylin and eosin (H&E) stained mouse brain slice. ANcr, cerebellar hemisphere ansiform lobule crus; arb, arbor vitae; CENT, cerebellar vermis central lobule; CUL, cerebellar vermis culmen; FL, cerebellar hemisphere flocculus; GRN, gigantocellular reticular nucleus; PFL, cerebellar hemisphere paraflocculus; SIM, cerebellar hemisphere simple lobule. **c**, Regions of interest (ROIs) denoted by the cyan rectangles in **a** and **b**. **d**, Profiles of the yellow dotted lines in **c**. **e**, ROIs denoted by the orange rectangles in **a** and **b**. **f**, Profiles of the yellow dotted lines in **e**. Norm., normalized. Scale bars, 200 μm.

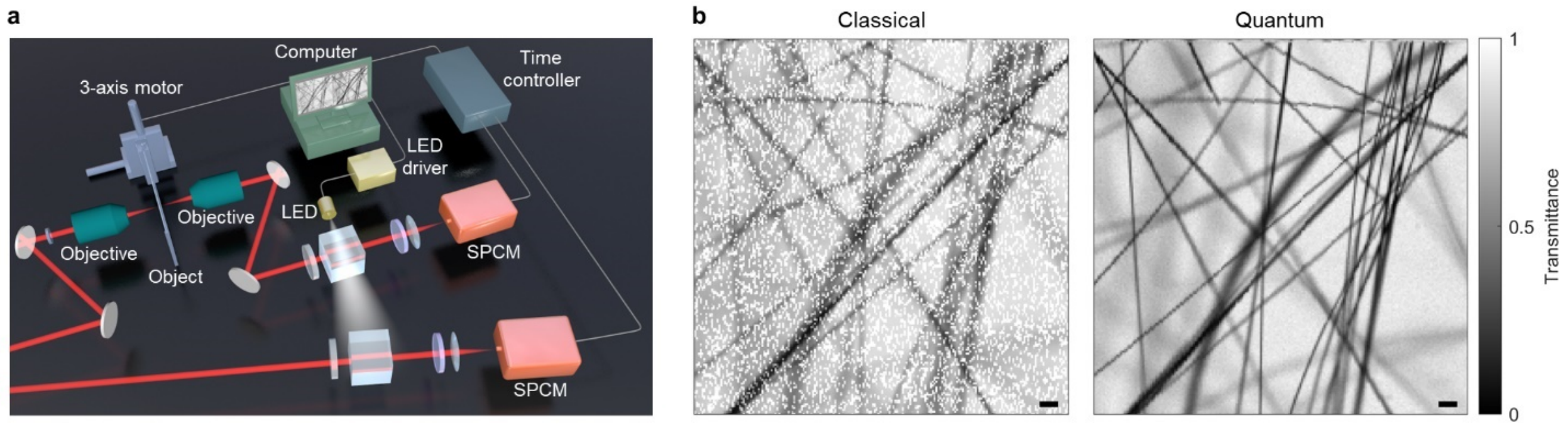

**Supplementary Fig. 10 ICE of carbon fibers in the presence of stray light.**

**a**, Experimental setup with the addition of a white light-emitting diode (LED) for randomly generated stray light. **b**, Classical and ICE images of carbon fibers in the presence of stray light. Norm., normalized. Scale bars, 100 μm.

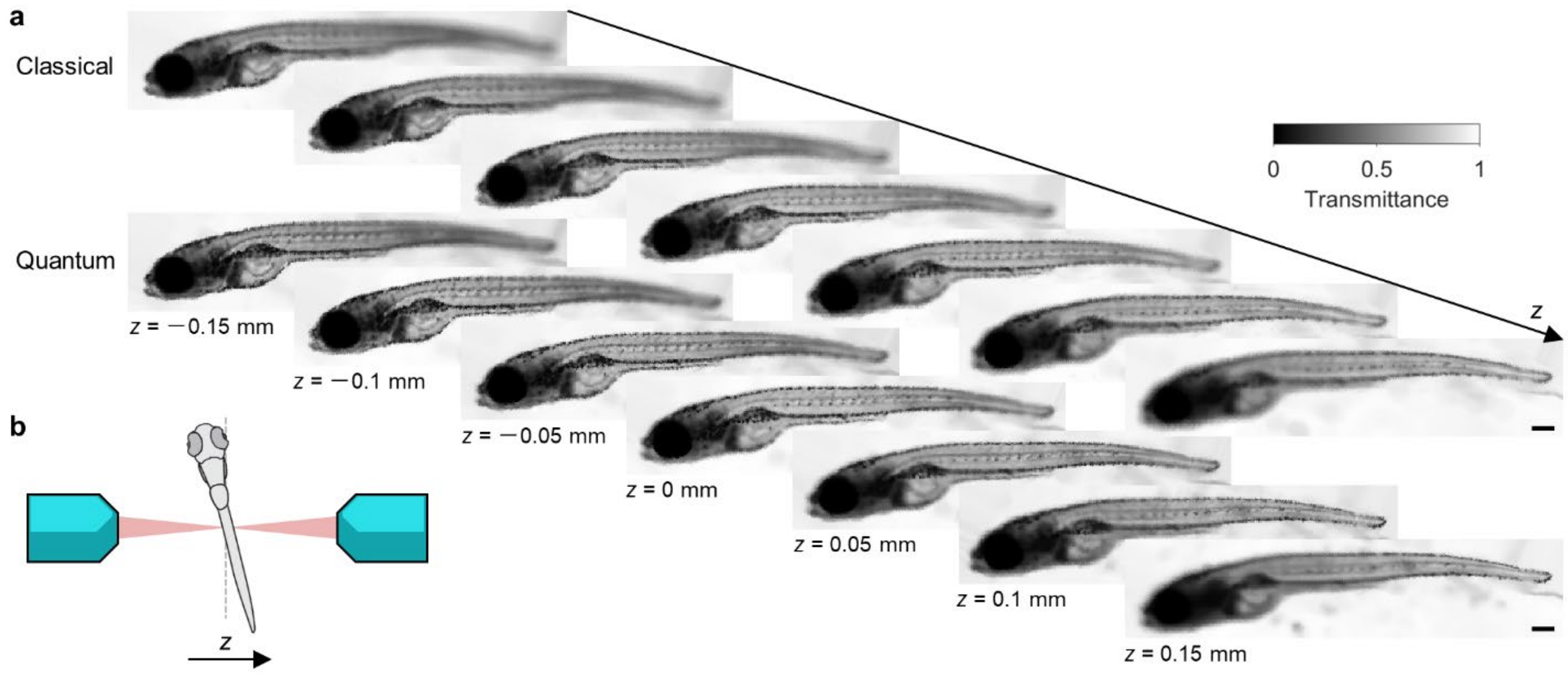

**Supplementary Fig. 11 ICE of a whole zebrafish.**

**a**, Classical and ICE images of an agarose-embedded zebrafish imaged at different *z* positions. Norm., normalized. Scale bars, 200 µm. **b**, Illustration of the imaging configuration, where the torso of the zebrafish is oblique to the imaging plane.

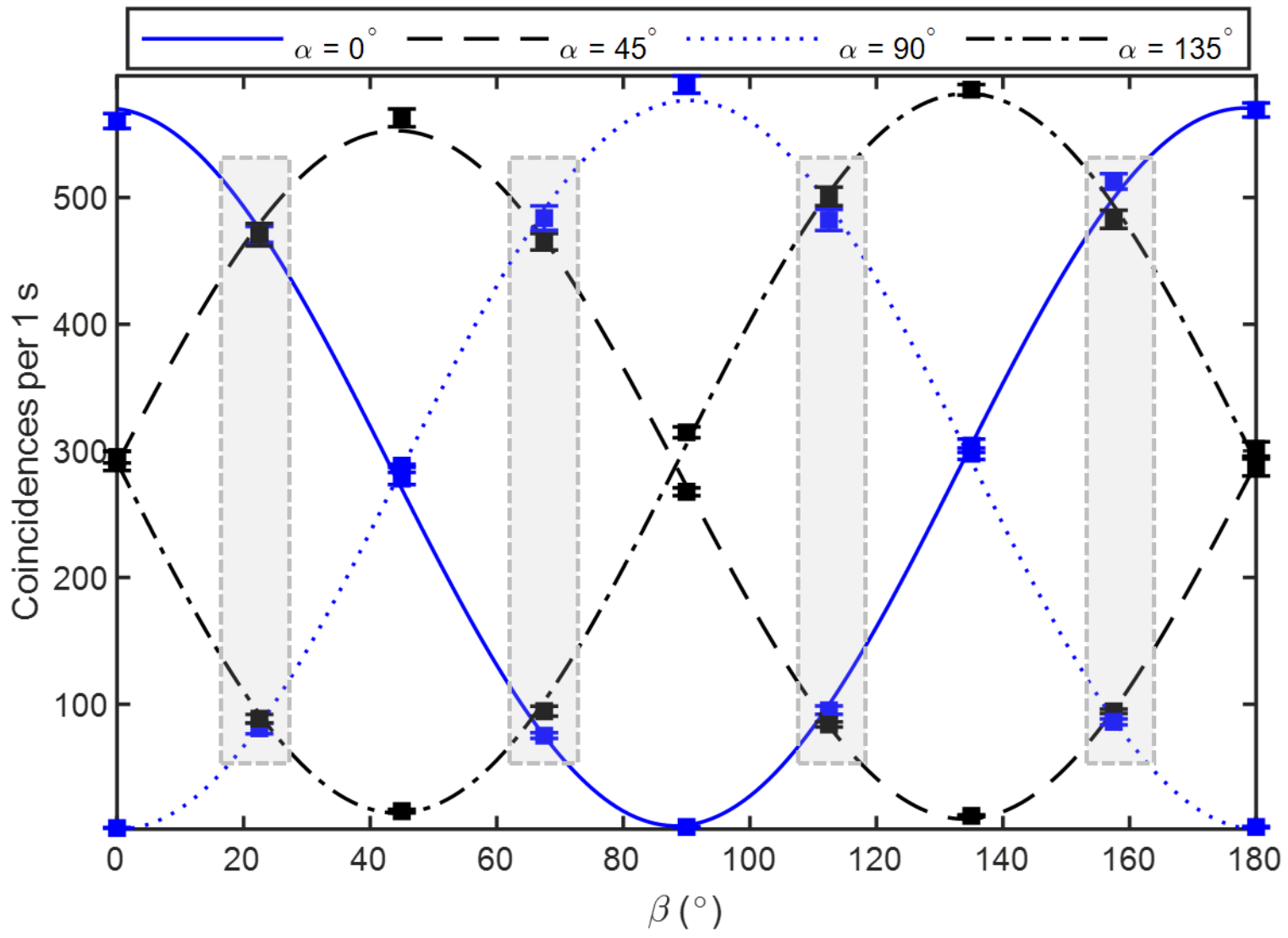

**Supplementary Fig. 12 Characterization of polarization entanglement through Bell's test.** Coincidence counts as a function of $\beta$ acquired with 1 s integration time for $\alpha = 0°, 45°, 90°, 135°$. Experimental results are plotted as means ± standard errors of the means. The points marked by gray dashed rectangles are used to calculate the $S_{\mathrm{CHSH}}$ value. The curves are fits based on Eq. (S38).

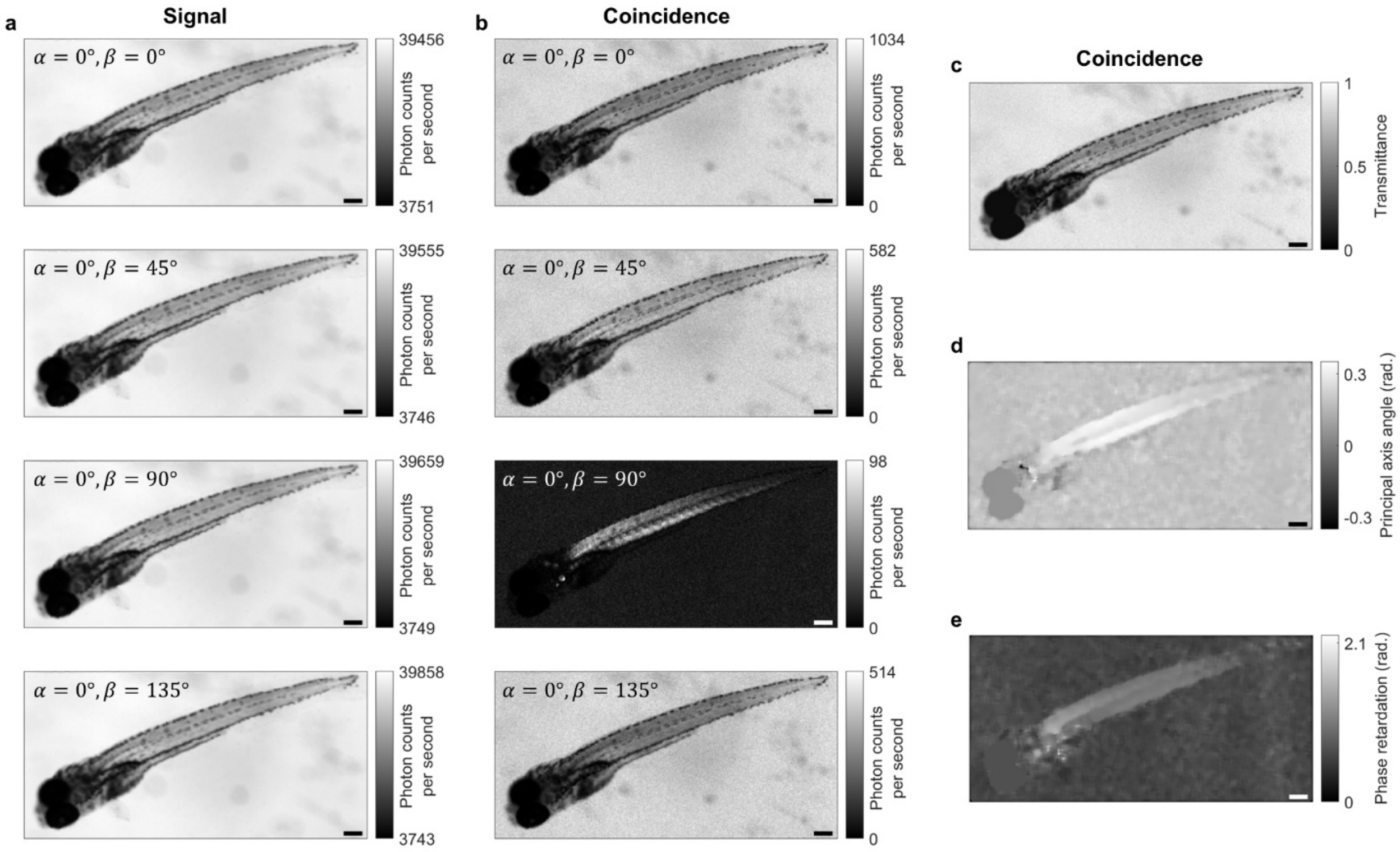

**Supplementary Fig. 13 Ghost birefringence imaging of a whole zebrafish using signal and coincidence counts.**

**a**, Classical images acquired with raw signal counts with constant $\alpha$ and variable $\beta$. **b**, ICE images acquired with coincidence counts with constant $\alpha$ and variable $\beta$. **c-e**, Transmittance (**c**), principal refractive index axis (**d**), and phase retardation between the two refractive index axes (**e**) calculated using the ICE images in **b**. Scale bars, 200 μm

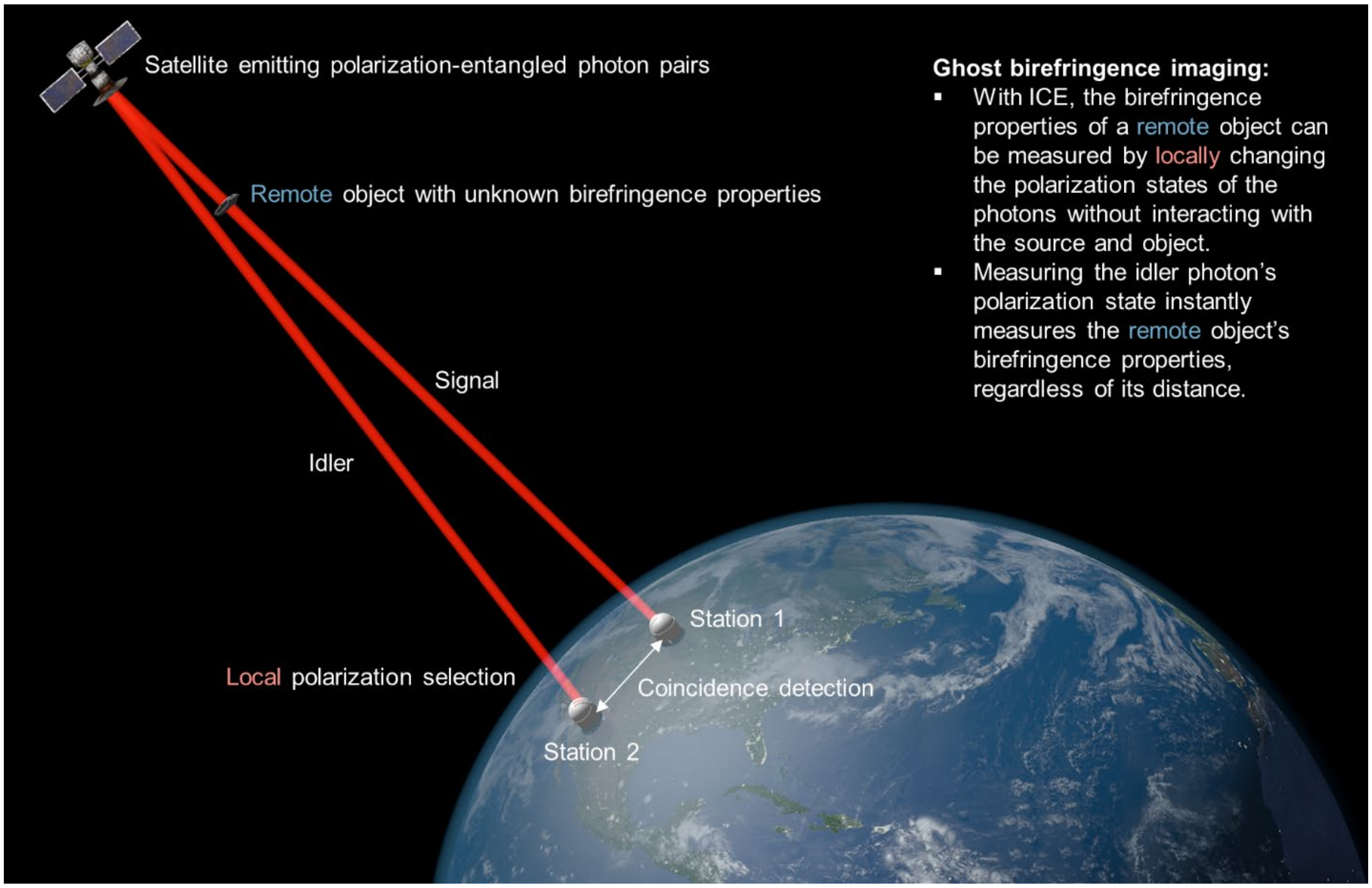

**Supplementary Fig. 14 Potential application of ghost birefringence imaging in remote sensing.**

With a satellite emitting polarization-entangled photon pairs[53,54], ICE can measure the birefringence properties of a remote object by changing the polarization states of the photons without interacting with the source and object. Through polarization entanglement, measuring the idler photon's polarization state instantly determines the incident signal photon's and, consequently, the remote object's birefringence properties, regardless of its distance.

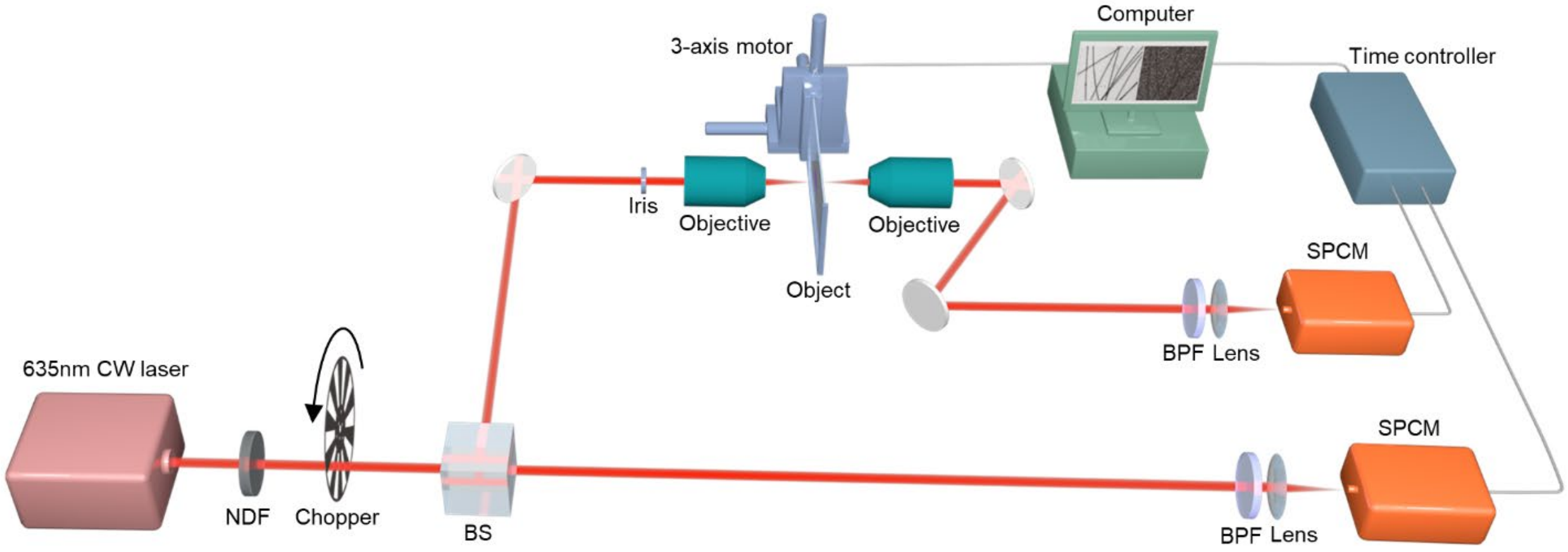

**Supplementary Fig. 15 Experimental setup of "ICE" with a classical source instead.**
CW, continuous wave; NDF, neutral density filter; BS, beam splitter; BPF, band-pass filter; SPCM, single-photon counting module.

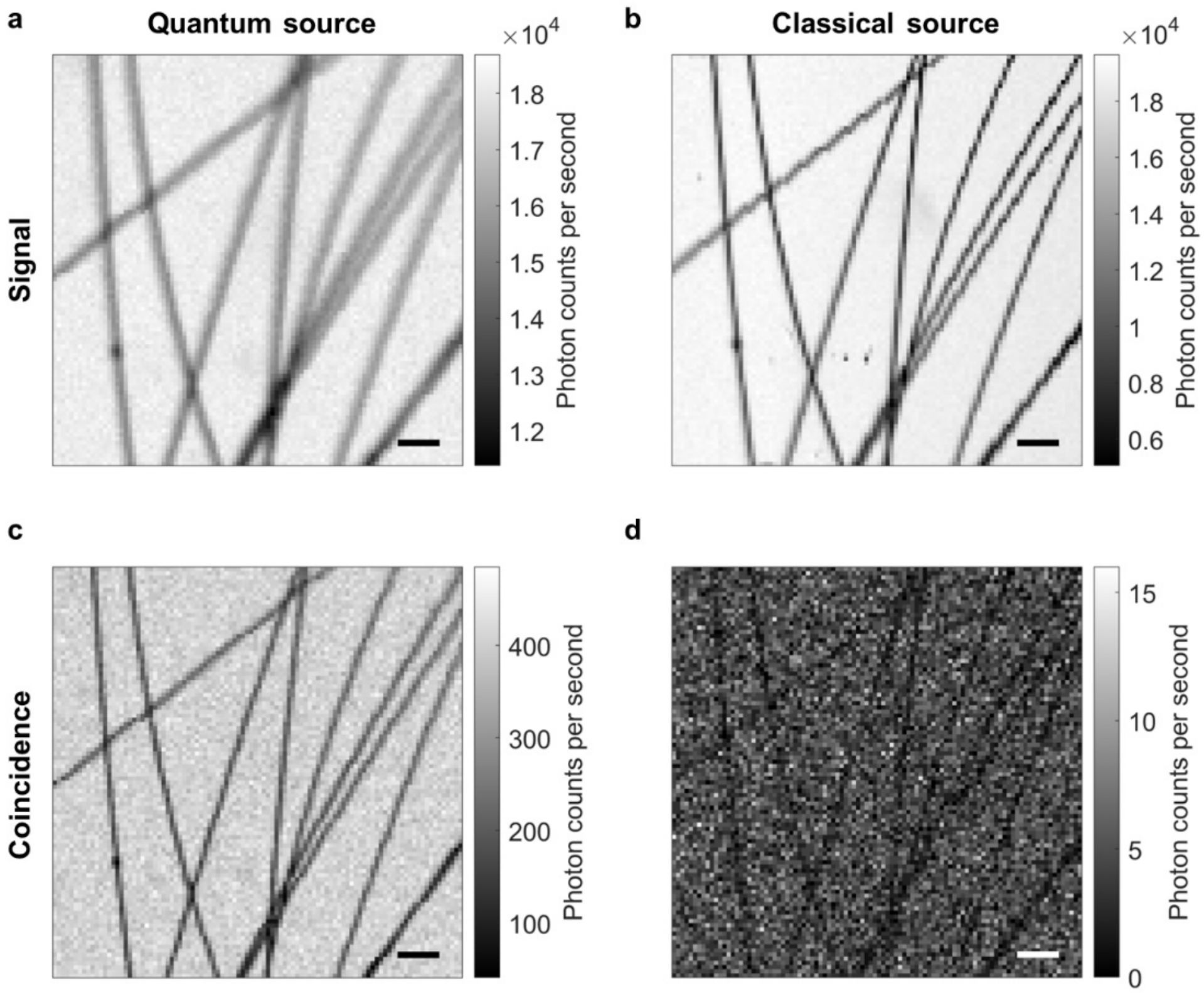

**Supplementary Fig. 16 ICE images of carbon fibers acquired using entangled and classical sources.**

**a**,**b**, Images formed with the raw signal counts using the entangled (**a**) and classical (**b**) sources. **c**,**d**, Images formed with the coincidence counts using the entangled (**c**) and classical (**d**) sources. Scale bars, 100 μm

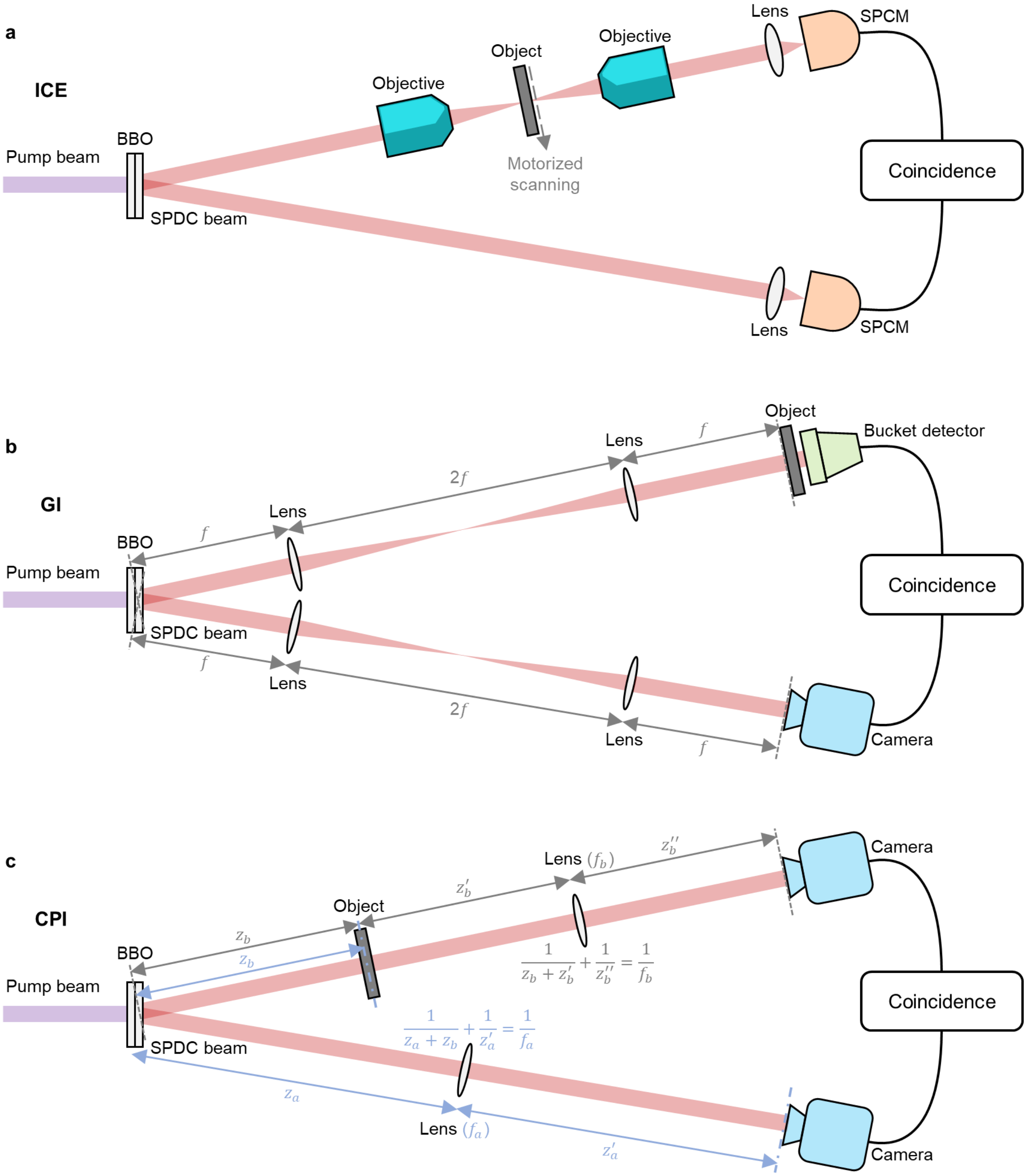

**Supplementary Fig. 17 Comparison of ICE, GI, and CPI.**

**a**, Schematic of ICE. The object is directly imaged to the SPCM in the signal arm. An image of the object is retrieved by recording the coincidence of the two SPCMs while raster scanning the object. **b**, Schematic of GI. The BBO crystal is imaged to both the object and the camera. The bucket detector records all the photons transmitted through the object. Through coincidence detection, a ghost image of the object is retrieved from the camera and triggered by the bucket detector. **c**, Schematic of CPI. The BBO crystal is imaged to the camera in the signal arm through the lens with a focal length of $f_b$, where the labeled distances satisfy the thin-lens equation $1/(z_b + z_b') + 1/z_b'' = 1/f_b$. The ghost image of the object is imaged to the camera in the idler arm though the lens with a focal length of $f_a$, satisfying the condition $1/(z_a + z_b) + 1/z_b' = 1/f_a$. Through coincidence detection, a ghost image of the object is retrieved from the camera in the idler arm, triggered by the camera in the signal arm.

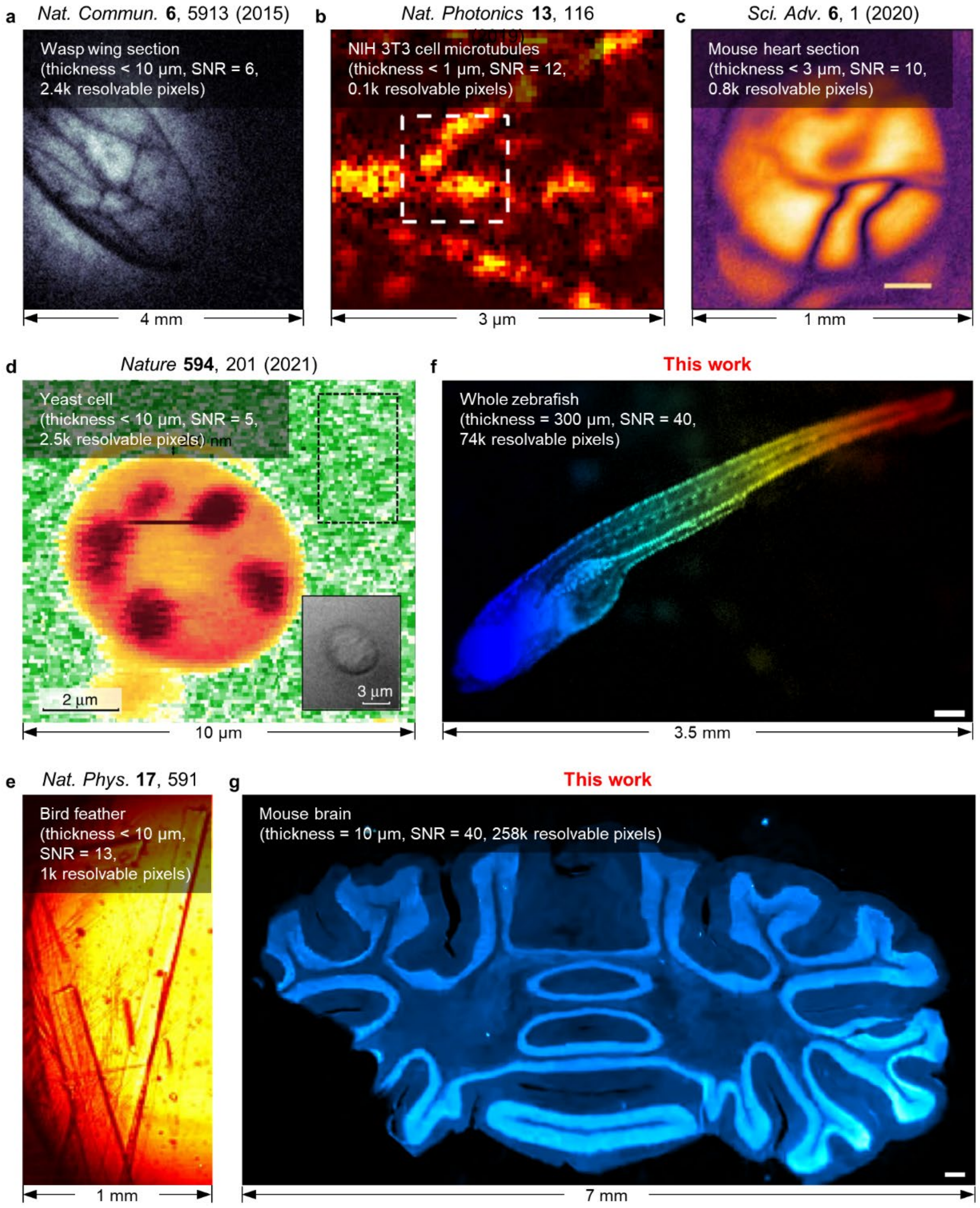

**Supplementary Fig. 18 Comparison of ICE and existing quantum bioimaging modalities.**
**a**, Quantum image of a wasp wing in a 4 mm × 4 mm FOV[7]. **b**, Quantum image of a 3 µm × 3 µm section of microtubules in a fixed 3T3 cell labeled with fluorescent quantum dots[8]. **c**, Quantum image of the histology sample of a mouse heart in a 1 mm × 1 mm FOV[9]. **d**, Quantum image of a yeast cell in aqueous buffer in a 10 µm × 10 µm FOV[11]. **e**, Quantum image of parts of a bird feather in a 2 mm × 1 mm FOV[10]. **f**,**g**, Quantum images of a whole zebrafish in a 3.5 mm × 2.3 mm FOV (**f**) and a mouse brain slice in a 7 mm × 4 mm FOV (**g**), presented in this work.

## Supplementary Table 1 Comparison of ICE and existing quantum bioimaging modalities

| Work | Specimen | Specimen thickness | FOV | Resolvable pixel count | Image formation | Stray light resilience | SNR | $\widehat{\mathrm{SNR}}$* ($s^{-0.5}$) |
|---|---|---|---|---|---|---|---|---|
| Ref. [7] | Wasp wing section | < 10 µm | 4 × 4 $mm^2$ | 2401 | Widefield | N/A | 6 | 7.2 |
| Ref. [8] | NIH 3T3 cell microtubules | < 1 µm | 3 × 3 $µm^2$ | 121 | Scanning | N/A | 12 | 9.5 |
| Ref. [9] | Mouse heart section | < 3 µm | 1 × 1 $mm^2$ | 784 | Widefield | N/A | 10 | 22.4 |
| Ref. [11] | Yeast cell | < 10 µm | 10 × 10 $µm^2$ | 2500 | Scanning | N/A | 5 | 11.2 |
| Ref. [10] | Bird feather | < 10 µm | 2 × 1 $mm^2$ | 968 | Widefield | Yes | 13 | 1.6 |
| **This work** | **Zebrafish / mouse brain** | **Up to 300 µm** | **Up to 7 × 4 $mm^2$** | **Up to 258432** | **Scanning** | **Yes** | **40** | **40** |

*Normalized SNR: $\widehat{\mathrm{SNR}} = \mathrm{SNR}/\sqrt{\text{Acquisition time per resolvable pixel}}$.